\documentclass[preprint,12pt,authoryear]{elsarticle}

\usepackage{amsmath,amssymb,bm}
\usepackage{graphicx}
\usepackage{pdfpages}
\usepackage{booktabs,array}
\newcolumntype{L}[1]{>{\raggedright\arraybackslash}p{#1}}
\usepackage[colorlinks=true,linkcolor=blue,citecolor=blue,urlcolor=blue]{hyperref}
\graphicspath{{figures/}}
\biboptions{round}

\newcommand{\ii}{\mathrm{i}}
\newcommand{\Uten}{\ensuremath{U_{10}}}
\newcommand{\ustar}{\ensuremath{u_*}}
\newcommand{\Sacc}{\ensuremath{S_{\mathrm{acc}}}}
\newcommand{\uS}{\ensuremath{\bm{u}_s}}

\journal{Ocean Modelling}

\begin{document}

\begin{frontmatter}

\title{Surface Stokes drift from compact drifting wave buoys}

\author[eodyn]{Alexey S. Mironov\corref{cor1}}
\ead{alexey.mironov@eodyn.com}
\author[odl]{Fabrice Collard}
\author[eodyn]{Gw\'ena\"ele Jan}
\author[ifremer]{Bertrand Chapron}
\cortext[cor1]{Corresponding author.}
\address[eodyn]{eOdyn, Technop\^ole Brest-Iroise, 115 rue Claude Chappe, 29280 Plouzan\'e, France}
\address[odl]{OceanDataLab, 870 route de D\'eolen, 29280 Locmaria-Plouzan\'e, France}
\address[ifremer]{Ifremer, Univ. Brest, CNRS, IRD, Laboratoire d'Oc\'eanographie Physique et
Spatiale (LOPS), ZI de la Pointe du Diable, CS 10070, 29280 Plouzan\'e, France}

\begin{abstract}
Surface Stokes drift depends strongly on the energy and directions of short waves, which are
incompletely resolved by routine wave observations. We derive surface Stokes drift vectors
from wave measurements collected by compact drifting buoys during three deployments in the
North-East Atlantic and the Alboran Sea. The calculation uses vertical-acceleration spectra and
first directional Fourier moments, which describe the mean wave direction and directional
concentration at each frequency; it accounts for the Doppler shift caused by buoy motion
relative to the water and adds a calibrated high-frequency tail above an intrinsic frequency
of $0.7$\,Hz. Across $13{,}139$ records, the median estimated speed is $0.081$\,m\,s$^{-1}$ at
a median wind speed of $6.8$\,m\,s$^{-1}$. Over the measured band of $0.04$--$1$\,Hz,
accounting for wave directions reduces the magnitude by a median $39\%$ relative to the
unidirectional assumption. The median ratio of the parameterised tail magnitude above $0.7$\,Hz to the total estimated
magnitude is $0.37$.
Comparisons with WAVEWATCH~III and Copernicus Marine MFWAM show strong covariation and similar
wind-dependent differences from the buoy-derived estimates. On the station-matched sample
from the two Atlantic deployments, WAVEWATCH~III directional spectra indicate that these differences within the compared band
arise mainly from spectral levels rather than from net directional reduction. The observations
provide constraints for model evaluation; the contribution of the unresolved short waves
remains sensitive to the assumed spectral tail and its directional spreading.
\end{abstract}


\end{frontmatter}

\section{Introduction}
\label{sec:intro}

The Stokes drift is the wave-induced difference between the Lagrangian-mean velocity of water
particles and the Eulerian-mean velocity measured at fixed positions \citep[][see \citealp{vandenbremer2018stokes}, for a
review]{stokes1847theory,longuethiggins1953mass,kenyon1969stokes,phillips1977dynamics}. At
the sea surface it is of the order of one per cent of the wind speed and contributes to the
transport of oil, plastic litter, search-and-rescue targets and the drifters used to observe
the ocean \citep{ardhuin2009observation,sutherland2020leeway,calvert2021litter,
pawlowicz2024drifters}. Its profile and shear enter the Coriolis--Stokes force and the
Langmuir-turbulence parameterisations of coupled wave--ocean models
\citep{polton2005stokescoriolis,breivik2016stokes}. These wave--current interactions depend
on the direction as well as the magnitude of the Stokes drift, so the quantity needed is the
Stokes drift vector rather than its speed alone.

Spectral wave models obtain the surface Stokes drift by integrating the frequency--direction
spectrum with a cubic frequency weighting, which increases the contribution of waves above
the spectral peak and makes the estimate sensitive to frequencies that are poorly resolved
or truncated in buoy observations \citep{rogers2025buoyhf,lenain2020hfstokes}. The short
waves that dominate the surface value are also the part of the spectrum where model spectra are
weakly constrained by observations: the model tail is parameterised, and its level and
directionality follow from source-term choices
\citep{lenain2020hfstokes,alday2023parameterizations}. Direct observations of the surface
Stokes drift vector remain rare, so modelled Stokes drift is seldom confronted with
measurements. A calculation based on the one-dimensional (omnidirectional) spectrum treats all waves as
travelling in one direction; its result, called here the unidirectional-equivalent magnitude,
overestimates the magnitude of the vector because differently directed wave components partly
cancel, a loss called here directional cancellation. For the records analysed here, waves
shorter than $10$\,m contribute about three fifths of the unidirectional-equivalent estimate
(Sect.~\ref{sec:sens}). The reduction due to spreading within each frequency is distinct
from that due to differently directed wave systems \citep{webb2015spreading}. Bulk
relations \citep{ardhuin2009observation,clarke2018relationship} do not retain the observed
frequency-dependent directional structure that spectral estimates carry \citep{kumar2017bulk}, and model-derived approximations
\citep{breivik2014approximate,breivik2016stokes} depend on the directional assumptions of
their source spectra. Buoy estimates of directional spreading carry their own limitations, in
particular for crossing seas \citep{lin2022spreading}, and drifting buoys observe the waves
at an encounter frequency, Doppler-shifted by their drift from the intrinsic one
\citep{rogers2026doppler}.

Directional cancellation and the importance of high-frequency waves are established
features of the surface Stokes drift. The observational difficulty is to distinguish
uncertainty in spectral level, directional structure and buoy response within the
frequencies that contribute most strongly. Compact drifting wave buoys provide
frequency-resolved directional moments along their trajectories, but still require a
continuation beyond their usable measurement band. Here we use MELODI, an expendable
undrogued drifter whose wind speed, friction velocity and wind-sea direction are retrieved
from the same records \citep{mironov2026windretrieval}, to address three questions. By how
much does the directional information reduce the buoy-derived magnitude relative to the
unidirectional-equivalent one? Do the differences from a wave model come mainly from this
directional reduction or from the spectral energy levels? How strongly do the assumptions
about the unresolved short waves affect the result? We estimate the surface Stokes drift
vector from wave spectra, not the separate contributions to the observed buoy motion. Supplementary material (S1--S12) provides derivations, processing and validation
details and exploratory analyses.

\section{Observations and estimation approach}
\label{sec:data}

\subsection{Primary observations and reference fields}
\label{sec:obs}

\begin{figure}[tbp]
  \centering
  \includegraphics[width=0.8\textwidth]{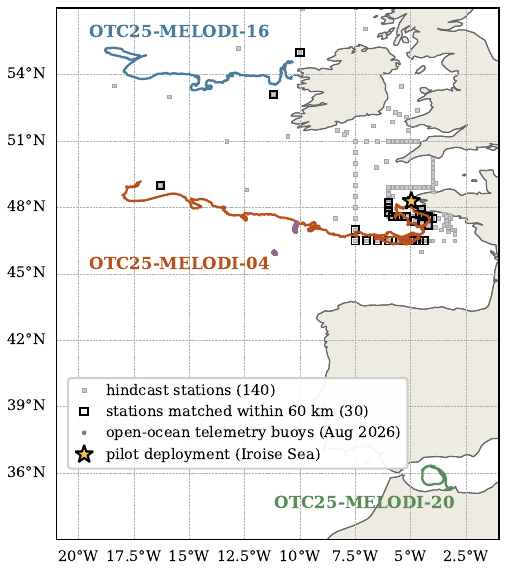}
  \caption{Tracks of the three primary deployments with model collocations, the pilot site
  (star), the two open-ocean telemetry buoys of August 2026 (dots) and the $140$ output
  stations of the GLOB-30M LOPS 2025 hindcast (small grey squares), of which $30$ are matched
  within $60$\,km of a record (black open squares). Map lines delineate the study area and do
  not necessarily depict accepted national boundaries.}
  \label{fig:tracks}
\end{figure}

The primary dataset comprises $13{,}139$ records from three recovered MELODI deployments
in the North-East Atlantic, including the Biscay and Iroise shelf and the storm track west of
Ireland, and in the Alboran Sea of the western Mediterranean (Table~\ref{tab:deploy},
Fig.~\ref{fig:tracks}). Each record is a $22$-min acquisition
repeated every $30$\,min. MELODI is a low-cost, expendable undrogued drifter developed by eOdyn
for sea-state monitoring \citep{mironov2023oceans,mironov2024igarss,mironov2026windretrieval}:
a disk hull of $240$\,mm diameter with a draft of about $5.5$\,cm, carrying a global
navigation satellite system (GNSS) receiver and a nine-axis inertial unit (S2). The GNSS receiver provides the positions; the wave
estimates of this paper use the inertial channels only. No independently validated correction
for the hull's wave-following response is applied (S9).

On the recovered buoys the attitude and the raw inertial channels are stored at $3.2$\,Hz.
From them the vertical-acceleration power spectral density $\Sacc(f)$, hereafter the
acceleration spectrum, and the normalised first directional Fourier coefficients
$a_1(f), b_1(f)$ are computed for each record from the cross-spectra of the gyroscope-derived
surface slopes with heave; the processing steps, moment definitions and record-selection
rules are listed in S3. Their complex combination $a_1 + \ii b_1$ describes the
energy-weighted mean wave direction and the directional concentration at each frequency: its
magnitude $r_1 = |a_1 + \ii b_1|$ is one for a unidirectional wave field and zero for an
isotropic field, although opposing wave components also give zero, and the first moments do
not determine the full directional distribution. Two inputs of the estimator are themselves
wave-derived. The wind speed $\Uten$, the equivalent neutral wind speed at $10$\,m height in
the scatterometer convention, is taken from the MELODI wind product, revision 1 of 27 August
2026, a spectral retrieval evaluated by leave-one-buoy-out validation against scatterometers
with a root-mean-square (RMS) difference of $0.90$\,m\,s$^{-1}$ \citep{mironov2026windretrieval}.
The air-side friction velocity $\ustar$ is inferred from the equilibrium level of $\Sacc(f)$ under the
equilibrium-range assumption of \citet{toba1973local} with a coefficient of $0.062$, and is
not a stress measurement. No external wind observations enter the estimate.

\begin{table}[tbp]
\centering\footnotesize
\caption{Primary deployments. The analysed period is the interval of the records with
per-frequency directional moments, all in 2025 (192, 71 and 13 days). The last column gives the number of records
with a collocated WAVEWATCH~III (WW3) field.}
\label{tab:deploy}
\begin{tabular}{L{0.22\textwidth}L{0.27\textwidth}L{0.15\textwidth}rr}
\toprule
deployment & region & analysed period & records & with WW3 \\
\midrule
\mbox{OTC25-MELODI-04} & North-East Atlantic, Biscay and Iroise shelf to $18^\circ$W, $47.2^\circ$N & 22 May--30 Nov & 9,129 & 9,105 \\
\mbox{OTC25-MELODI-16} & North Atlantic storm track, $54.0^\circ$N & 14 May--24 Jul & 3,384 & 3,384 \\
\mbox{OTC25-MELODI-20} & Alboran Sea, western Mediterranean, $35.5^\circ$N & 27 May--9 Jun & 626 & 626 \\
\midrule
all & winds $0$--$18$\,m\,s$^{-1}$, $H_s$ $0.3$--$6$\,m & & 13,139 & 13,115 \\
\bottomrule
\end{tabular}
\end{table}

Three supporting datasets serve specific checks. High-rate ($100$\,Hz) inertial records from
a pilot deployment of four buoys in the Iroise Sea provide a check of the spectral slope up to
$1$\,Hz (S11). Two open-ocean drifters of the August 2026 firmware, deployed west of Biscay,
provide directional comparisons below $0.5$\,Hz (S11). The fleet's archive of $8{,}245$
telemetered omnidirectional acceleration spectra from March 2025 to March 2026 supports a separate approximation
for records without frequency-resolved directional moments (S2, S8).

\begin{sloppypar}
The buoy-derived estimates are compared with two model products, WAVEWATCH~III in two
configurations and MFWAM, whose versions, grids, spectral supports and collocation rules are
summarised in Table~S8 (S12).
\end{sloppypar}

Gridded surface Stokes drift fields of WAVEWATCH~III (WW3), a third-generation spectral wave
model \citep{ww3dg2019manual}, are taken from the Ifremer/LOPS hindcast
GLOBMULTI\_\allowbreak ERA5\_\allowbreak GLOBCUR\_01 \citep{accensi2020globmulti,alday2021database}, forced by ERA5
winds and CMEMS GLOBCURRENT currents, on its North-East Atlantic $10'$ grid with the global
$0.5^\circ$ grid where the regional grid is unavailable. The 3-hourly fields are interpolated
bilinearly in space and linearly in time to the buoy positions and matched to the records
($N = 13{,}115$). The distributed WW3 Stokes-drift diagnostic integrates the prognostic
directional spectrum over the model's $36$ frequencies, the last bin centred at $0.95$\,Hz,
so it includes the model's own directional reduction and omits an analytical high-frequency
tail (S12); the buoy-derived estimate, by contrast, includes an extrapolated tail.

For the spectrum-to-spectrum comparison, directional spectra are taken from the Ifremer/LOPS
hindcast GLOB-30M\_\allowbreak LOPS\_2025 \citep{accensi2024glob30m}, a global $0.5^\circ$ run with the
parameterisation of \citet{alday2023parameterizations}, at its $140$ fixed output stations
within the track envelopes ($36$ frequencies to $0.95$\,Hz, $24$ directions, 3-hourly). They
are reduced with the same integrals as the buoy spectra and matched to records within
$60$\,km and $90$\,min: $4{,}610$ records at $30$ stations, median distance $20$\,km, of which
$4{,}383$ are from OTC25-MELODI-04, $227$ from OTC25-MELODI-16 and none from the Alboran Sea
deployment. The station run differs from the gridded hindcast in version, grid and forcing,
so the station comparison describes this subset and is not a decomposition of the
gridded-field differences.

In parallel, the surface Stokes drift of the Copernicus Marine Service global wave analysis,
produced with the MFWAM model of M\'et\'eo-France at $1/12^\circ$ and 3-hourly
\citep{cmems2025wav027}, is collocated bilinearly in space to the nearest analysis within
$90$\,min ($N = 12{,}889$, the Alboran records nearest the coast having no valid bilinear
interpolation). MFWAM integrates its spectrum to $0.58$\,Hz and, in the public ECWAM routine on which
the operational code is based, adds the analytic contribution of an $f^{-5}$ tail matched to
the spectral level there (S12), so its Stokes
drift also includes an extrapolated tail, with a shape and level that differ from the buoy's
calibrated tail. ERA5 winds serve as a directional reference and the bulk relation of
\citet{ardhuin2009observation} in the buoy-derived wind speed and wave height as a benchmark
(S12). Wave and drift directions are propagation-toward bearings clockwise from north; in
the equations the wave propagation direction $\theta$ is measured counterclockwise from east. Successive records, storms
and model fields are correlated, so sampling uncertainties are $95\%$ intervals from a block
bootstrap that resamples 3-day blocks within each deployment (replicate counts in S5). These
intervals describe sampling uncertainty, not the uncertainty from the assumed high-frequency
tail or from the instrument response.

\subsection{The estimator}
\label{sec:estimator}

\begin{figure}[tbp]
  \centering
  \includegraphics[width=\textwidth]{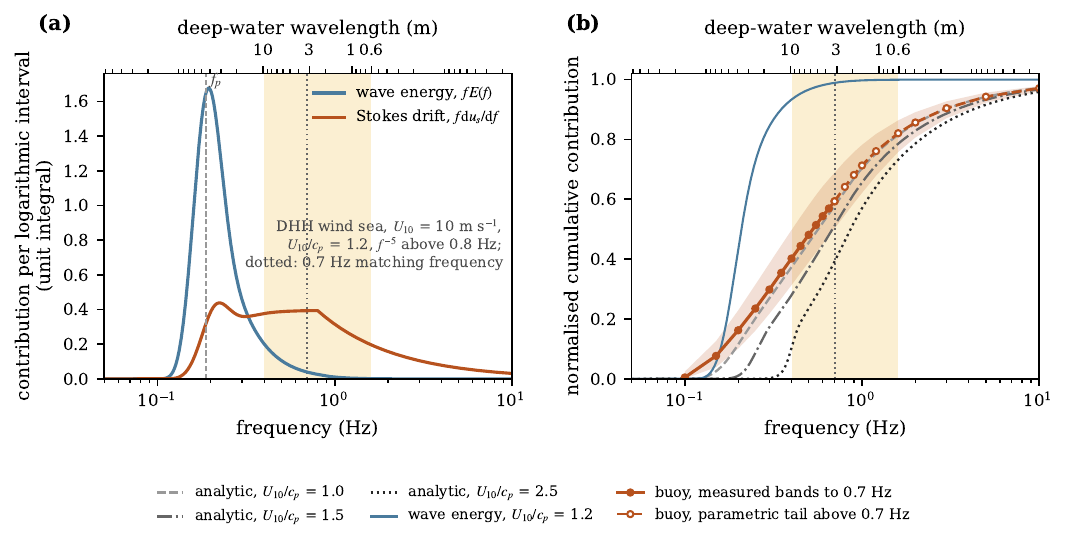}
  \caption{Frequency distribution of wave energy and of the unidirectional-equivalent surface
  Stokes drift. (a) Contributions per logarithmic frequency interval, $f\,E(f)$ and
  $f\,\mathrm{d}u_s/\mathrm{d}f$, normalised to unit integral, for an analytic Donelan--Hamilton--Hui (DHH) wind sea
  \citep{donelan1985directional} ($\Uten = 10$\,m\,s$^{-1}$, $\Uten/c_p = 1.2$ with $c_p$ the
  phase speed at the spectral peak, $f^{-5}$ saturation above $0.8$\,Hz).
  (b) Normalised cumulative contribution: analytic wind seas at three inverse wave ages, the
  wave energy of the $\Uten/c_p = 1.2$ sea, and the buoy-derived reference estimate without
  directional reduction as median and interquartile range over the $13{,}139$ records, measured bands to the
  matching frequency of $0.7$\,Hz (filled symbols) and the parametric tail beyond (open
  symbols). The shaded band is $0.4$--$1.6$\,Hz, wavelengths $10$\,m to $0.6$\,m; the dotted
  line the matching frequency. The cumulative curves are unidirectional-equivalent estimates,
  not measured full-spectrum vectors.}
  \label{fig:weight}
\end{figure}

In deep water the surface Stokes drift of a directional spectrum is the Kenyon integral
\citep{kenyon1969stokes}, which with the omnidirectional elevation spectrum $E(f)$, a one-sided density in
m$^2$\,Hz$^{-1}$, and the first directional coefficients defined by $\int E(f,\theta)e^{\ii\theta}d\theta = E(f)[a_1 + \ii b_1]$
can be written in complex notation, the real and imaginary parts being the eastward and
northward components,
\begin{equation}
  \uS \;=\; \frac{16\pi^3}{g}\int_0^\infty f^3 E\,\bigl[a_1+\ii\,b_1\bigr]\, df
  \;=\; \frac{1}{\pi g}\int_0^\infty \frac{\Sacc}{f}\,\bigl[a_1+\ii\,b_1\bigr]\, df ,
  \label{eq:kenyon}
\end{equation}
where $\Sacc(f) = (2\pi f)^4 E(f)$ is the one-sided vertical-acceleration spectrum in
(m\,s$^{-2}$)$^2$\,Hz$^{-1}$, $g$ the acceleration due to gravity and $\theta$ the propagation
direction. The first directional moments are the only directional information the integral
needs. Deep-water dispersion is assumed for all records, including those over the shelf,
which were not screened by depth; for idealised wind-sea spectra with a swell partition the
finite-depth surface integral exceeds the deep-water one by a factor of at most $1.04$ at
$50$\,m depth and $1.12$ at $30$\,m (S1). The acceleration form is
algebraically identical to the elevation form, including its noise, and makes explicit that
an additive acceleration-noise floor enters with a $1/f$ weighting (S1). The $f^3$ weighting follows from the dispersion relation and makes the
estimate sensitive to high-frequency spectral levels, directional structure and instrument
response. In an $f^{-4}$ equilibrium range equal logarithmic frequency intervals contribute
equally to the unidirectional-equivalent integral, and the contribution decreases through an
$f^{-5}$ saturation range. Figure~\ref{fig:weight} contrasts this weighting with that of the
wave variance and identifies the observed and extrapolated parts of the estimate: the
calculation requires directional information within the measured band and explicit
assumptions beyond it.

The estimator sums \eqref{eq:kenyon} over the resolved bands and adds a parameterised
high-frequency tail; with the settings given below its result is the reference estimate of
this paper,
\begin{equation}
  \hat{\uS} \;=\; \frac{1}{\pi g}\sum_{f_{i,j} \le f_c}
      \frac{\Sacc(f_j)}{f_j}\Bigl(\frac{f_{i,j}}{f_j}\Bigr)^{3}(a_{1,j}+\ii\, b_{1,j})\,\Delta f_j
  \;+\; \Delta u_s^{\mathrm{hf}}(f_c)\, r_1^{\mathrm{ws}}\, e^{\ii\theta^{\mathrm{ws}}} .
  \label{eq:estimator}
\end{equation}
Here $f_j$ and $\Delta f_j$ are the measured centre frequency and width of band $j$, the sum
starts at $0.04$\,Hz, below which the moments are not reliable, and $f_{i,j}$ is the
intrinsic frequency of the band; $f_c$ is the matching frequency at which the resolved sum ends
and the parameterised tail begins. The first term is the resolved vector $\bm{u}_{\mathrm{res}}$
and the second the tail vector $\bm{u}_{\mathrm{hf}}$; their magnitudes are $u_{\mathrm{res}}$ and
$u_{\mathrm{hf}}$, the speed of the estimate is $\hat u_s = |\hat{\uS}|$, and the tail-to-total
magnitude ratio, called the tail ratio below, is $\Phi_{\mathrm{hf}} = u_{\mathrm{hf}}/\hat u_s$. It is not
an additive share of the speed: it exceeds one in $157$ records in which the resolved and tail
vectors partly cancel. A drifting buoy observes the encounter frequency
$2\pi f_e = 2\pi f_i + \bm{k}\cdot(\bm{U} - \bm{V}_b)$, with $\bm{k}$ the wavenumber vector,
$\bm{U}$ the current advecting the wave phase and $\bm{V}_b$ the buoy velocity
\citep{longuethiggins1986eulerian,rogers2025buoyhf}. The encounter frequency $f_e$ is the
frequency observed by the moving buoy and the intrinsic frequency $f_i$ that in a frame moving
with the mean current; their difference, a steady shift by the buoy's mean velocity relative
to the water, is called here the drift Doppler shift. It is distinct from the orbital
modulation of short waves by long waves, not treated here, and the target is the
intrinsic-frequency integral rather than the fixed-frame spectrum of
\citet{rogers2026doppler}. Because the acceleration is recorded in encounter time, the integrand written in
encounter coordinates acquires the factor $(f_i/f_e)^3$, and the resolved sum ends where the
intrinsic frequency reaches the matching frequency $f_c$, so that the resolved part and the
tail are defined on one intrinsic frequency axis. For this correction the buoy velocity
relative to the water, $\bm{V}_b - \bm{U}$, is approximated by $0.019\,\Uten$ in the direction
of the record's Stokes drift, the wind-coherent drift of the hull on the same records (S4). This is an assumed relative
velocity for the frequency transformation, not a measured windage or Stokes contribution to
the buoy motion; its magnitude and direction are varied in the sensitivity tests. The drift-Doppler factor $D_{\mathrm{d}} = u_{\mathrm{res}}/u_{\mathrm{res}}^{(e)}$, the
ratio of the resolved magnitude to its value computed on encounter frequencies, is $1.09$ at
the median, from $1.06$ in winds below $6$\,m\,s$^{-1}$ to $1.14$ above $14$\,m\,s$^{-1}$; the
correction rotates the resolved vector by $0.2^\circ$. The reference
estimate uses $f_c = 0.7$\,Hz. Table~\ref{tab:bands} lists the frequency bands and frames used
in this paper. The analysis band, $0.04$--$1$\,Hz, over which the directional reductions
defined at the end of this section are evaluated, extends beyond $f_c$: the measured
$0.7$--$1$\,Hz interval serves the directional diagnostics and the model comparison but is
deliberately excluded from the resolved part of the reference estimate.

\begin{table}[tbp]
\centering\footnotesize
\caption{Frequency bands and frames. The intrinsic frame is the mean-current frame; the
encounter frame is the frequency observed by the drifting buoy (Sect.~\ref{sec:estimator}).}
\label{tab:bands}
\begin{tabular}{L{0.17\textwidth}L{0.12\textwidth}L{0.60\textwidth}}
\toprule
band (Hz) & frame & use \\
\midrule
$0.04$--$0.7$ & intrinsic & measured contribution to the reference estimate (Eq.~\ref{eq:estimator}) \\
above $0.7$ & intrinsic & parameterised tail of the reference estimate \\
$0.6$--$0.7$ & intrinsic & calibration band of the tail constants (S5) \\
$0.04$--$1.0$ & intrinsic & analysis band: directional reductions $R$, $R_{\mathrm{sp}}$, $R_{\mathrm{md}}$ and the comparison with the model spectra (model bins to $0.95$\,Hz) \\
$0.60$--$0.90$ & encounter & direction and concentration of the tail; Stokes-weighted $\langle r_1\rangle$ diagnostics \\
bins of $f/f_p$ and $0.1$-Hz bands & encounter & frequency-resolved concentration $r_1$ (Fig.~\ref{fig:r1_ffp}) \\
$0.275$--$0.475$ & encounter & friction velocity from the equilibrium level of $\Sacc$ \\
\bottomrule
\end{tabular}
\end{table}

The matching frequency is set at $0.7$\,Hz because the measurement response becomes
uncertain at higher frequencies. The signal exceeds the electronic floor by five to six orders of magnitude there, but above
$0.7$\,Hz the elevation spectrum inferred from the $3.2$\,Hz records steepens where the
$100$\,Hz pilot accelerometer and the model tail do not, and no independent validation of the
channel's high-frequency response is available (Sect.~\ref{sec:ww3}, S11). Above $f_c$ the
spectrum is continued with an equilibrium range $E = \beta_T\, g\, \ustar\, (2\pi)^{-3}
f^{-4}$ \citep{toba1973local,phillips1985equilibrium} up to the transition
$f_t = \alpha_P g/(2\pi\beta_T\ustar)$ and a saturation range
$E = \alpha_P\, g^2\, (2\pi)^{-4} f^{-5}$ above it
\citep{phillips1985equilibrium,breivik2016stokes}. Both integrate in closed form to the
unidirectional magnitude $\Delta u_s^{\mathrm{hf}}(f_c)$ (S5). The direction
$\theta^{\mathrm{ws}}$ and concentration $r_1^{\mathrm{ws}}$ of the tail are the argument and
magnitude of the acceleration-spectrum-weighted complex first moment
$\sum_j \Sacc(f_j)(a_{1,j}+\ii b_{1,j})/\sum_j \Sacc(f_j)$ over the measured
$0.60$--$0.90$\,Hz band.

The two constants are effective tail parameters for this sensor, fitted by directional
closure: the parametric $0.6$--$0.7$\,Hz contribution, multiplied by the tail concentration
$r_1^{\mathrm{ws}}$, is matched to the measured $0.6$--$0.7$\,Hz intrinsic-band contribution
projected onto the tail direction $\theta^{\mathrm{ws}}$, through the median closure ratio in
each of four wind classes ($\Uten$ below $6$, $6$--$10$, $10$--$14$ and above $14$\,m\,s$^{-1}$,
used throughout the paper; S5). The fit gives $\beta_T = 0.069$ and
$\alpha_P = 1.32\times10^{-2}$; the fitted values absorb the prescribed tail direction and
concentration as well as the spectral level and the sensor and processing response, and are
not independent estimates of universal equilibrium-range constants. Fitting details,
bootstrap intervals and the comparison with published constants are given in S5. Because $\ustar$ is itself inferred
from a lower-frequency band, $\beta_T$ partly adjusts the relative spectral levels between
bands. Calibration stability is assessed by fitting the constants to two deployments and
applying them to the third: on the held-out records, the ratio of the median speed with the held-out constants to that
with the all-data constants lies within $0.98$--$1.01$ (S5). This
tests the sensitivity to the calibration sample; it does not independently validate the wave
spectrum above $0.7$\,Hz, so the unresolved contribution is assessed through its sensitivity
to the assumed tail energy and directional spreading (Sect.~\ref{sec:sens}). The tail
direction and concentration still come from the $0.60$--$0.90$\,Hz encounter band, so the
matching frequency does not remove the sensitivity to differential channel response in that
band.

The directional moments use timing-corrected slope--heave cross-spectra. Synthetic seas with
the measured electronic noise floor recover the prescribed moments under the tested
conditions (S3), but no independent in situ measurement of the high-frequency moments is
available, so they remain a measurement limitation rather than an independently validated
quantity; their comparison with a published parameterisation is given in
Sect.~\ref{sec:directional}. The sensitivity of the estimate to the tail
constants, the matching frequency, the closure band, the directional structure of the tail
and the assumptions of the drift-Doppler correction was tested by recomputing the affected
part of the estimate for each alternative with the other settings at their reference values;
for the drift-Doppler alternatives the resolved contribution is recomputed and the reference
tail is retained, without a new fit of the tail constants or moments (S4--S6). The outcomes
are reported in Sect.~\ref{sec:sens}. Alternatives and model products are compared with the
reference estimate through three ratios,
\begin{equation}
  \tilde\rho_x = \frac{\operatorname{med}\,\hat u_s^{(x)}}{\operatorname{med}\,\hat u_s}, \qquad
  \rho_x = \operatorname{med}\,\frac{\hat u_s^{(x)}}{\hat u_s}, \qquad
  \rho_{\mathrm{m}} = \frac{\overline{u_{\mathrm{m}}}}{\overline{\hat u_s}} ,
  \label{eq:ratios}
\end{equation}
where $\hat u_s^{(x)}$ is the speed obtained with alternative $x$, $\operatorname{med}$ the sample
median over records, $u_{\mathrm{m}}$ the speed of a model product and the overbar the mean over
the matched records. The sample-median ratio $\tilde\rho$ is used when the alternative changes
the calibration of the estimator (tail constants, matching frequency, bootstrap refits), the
median recordwise ratio $\rho$ when it changes the estimate record by record with the reference
constants, and the mean-speed ratio $\rho_{\mathrm{m}}$, written $\rho_{\mathrm{WW3}}$ and
$\rho_{\mathrm{MF}}$ for the two products, for the model comparison, over all matched records
or within a wind class.

If all waves are assumed to travel in one direction, their Stokes-drift contributions add
without cancellation; accounting for their directions reduces the magnitude of the resulting
vector. We quantify this by $R$, the fraction of the unidirectional-equivalent magnitude
retained, following \citet{webb2015spreading}:
\begin{equation}
\begin{gathered}
  u_{\mathrm{1D}} = \frac{16\pi^3}{g}\!\int f^3 E\,df, \qquad
  u_{\mathrm{2D}} = \Bigl|\frac{16\pi^3}{g}\!\int f^3 E\,(a_1{+}\ii b_1)\,df\Bigr|, \\
  R \equiv \frac{u_{\mathrm{2D}}}{u_{\mathrm{1D}}} = R_{\mathrm{sp}}\,R_{\mathrm{md}},
\end{gathered}
\label{eq:reduction}
\end{equation}
where $R_{\mathrm{sp}} = \langle r_1\rangle$, the mean of $r_1$ weighted by each band's contribution
to $u_{\mathrm{1D}}$ (Stokes-weighted), is the fraction retained after accounting for directional spreading within frequencies, and
$R_{\mathrm{md}}$ the additional fraction retained after combining frequency bands with
different mean propagation directions, as can occur between wind sea and swell; the
fractional reduction is $1 - R$. The identity holds record by record. The same Stokes-weighted
mean over a restricted band is written with the band as subscript, for example
$\langle r_1\rangle_{0.4\text{--}0.7}$. These directional statistics are evaluated on the analysis
band ($0.04$--$1$\,Hz), with the same intrinsic selection as the estimator.

\section{Results}
\label{sec:results}

\subsection{Effects of wave direction and spectral contributions}
\label{sec:directional}

Accounting for wave directions substantially reduces the estimated surface Stokes-drift
speed. Over the analysis band ($0.04$--$1$\,Hz) the median retained fraction $R$ is $0.61$ on
the $13{,}115$ model-collocated records ($95\%$ interval $0.60$--$0.61$). Most of this loss arises between waves of
the same frequency travelling in different directions (median $R_{\mathrm{sp}} = 0.66$), and
little between differently directed frequency bands ($R_{\mathrm{md}} = 0.94$). $R$ varies little with wind, from $0.59$ below $6$\,m\,s$^{-1}$ to
$0.54$ above $14$\,m\,s$^{-1}$, while the concentration in the $0.60$--$0.90$\,Hz band falls
from $0.64$ to $0.40$. Leaving out any one deployment moves $R$ within $0.59$--$0.61$.
Because $R_{\mathrm{sp}}$ is the Stokes-weighted mean of $r_1$, which frequencies contribute
most to the cancellation depends on how $r_1$ varies across the band that carries the drift.

\begin{figure}[tbp]
  \centering
  \includegraphics[width=\textwidth]{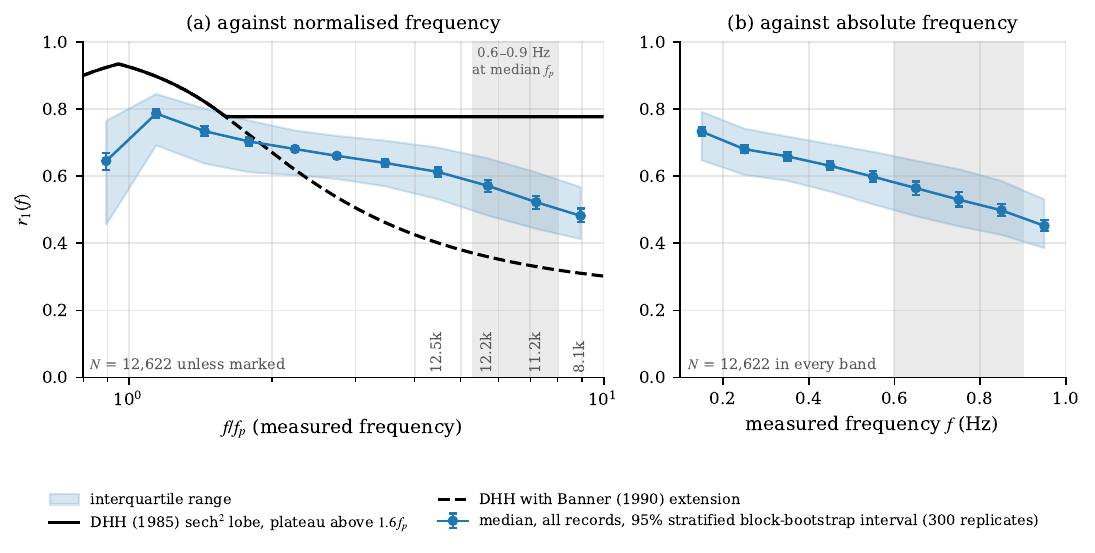}
  \caption{Directional concentration $r_1$ of the recovered-buoy records ($N = 12{,}622$,
  measured frequencies): median with its $95\%$ block-bootstrap interval ($300$ replicates
  of 3-day blocks within deployments) and interquartile range, (a) against $f/f_p$ with the
  DHH $\mathrm{sech}^2$ lobe and its plateau above $1.6 f_p$ and the Banner (1990) extension,
  both normalised over all directions, and (b) against absolute frequency. The grey band marks
  $0.6$--$0.9$\,Hz, at the median $f_p$ of $0.11$\,Hz in (a). The coherence diagnostic and
  the high-peak-frequency subset are shown in S3.}
  \label{fig:r1_ffp}
\end{figure}

The directional concentration $r_1$ decreases steadily with frequency above the peak, so
shorter waves have less concentrated propagation directions (Fig.~\ref{fig:r1_ffp}). Against the frequency relative to the
spectral peak, $f/f_p$, the median $r_1$ peaks at $0.79$ near the peak, $0.07$--$0.11$ below the
concentration predicted by the Donelan--Hamilton--Hui (DHH) form \citep{donelan1985directional}
between $1$ and $2 f_p$ ($12{,}622$ records with $0.06 \le f_p \le 0.4$\,Hz, measured frequencies;
$95\%$ block-bootstrap half-widths of $0.01$--$0.02$ on the bin medians). Above about $2 f_p$ it lies between
the two predicted concentrations, the constant DHH value of $0.78$ and the continued
broadening of \citet{banner1990equilibrium}, decreasing to $0.64$ at $3$--$4 f_p$ and $0.48$
at $8$--$10 f_p$, an intermediate concentration in the band that sets $R_{\mathrm{sp}}$. In
absolute frequency the median decreases from $0.73$ at $0.1$--$0.2$\,Hz to $0.45$ at
$0.9$--$1.0$\,Hz. The two panels do not separate sea-state dependence from instrument
response. Directional moments from the separate open-ocean deployments show a similar trend
below $0.5$\,Hz (S11), and agreement of an integrated moment does not establish equality of
the directional distributions \citep{lin2022spreading}.

\begin{figure}[tbp]
  \centering
  \includegraphics[width=\textwidth]{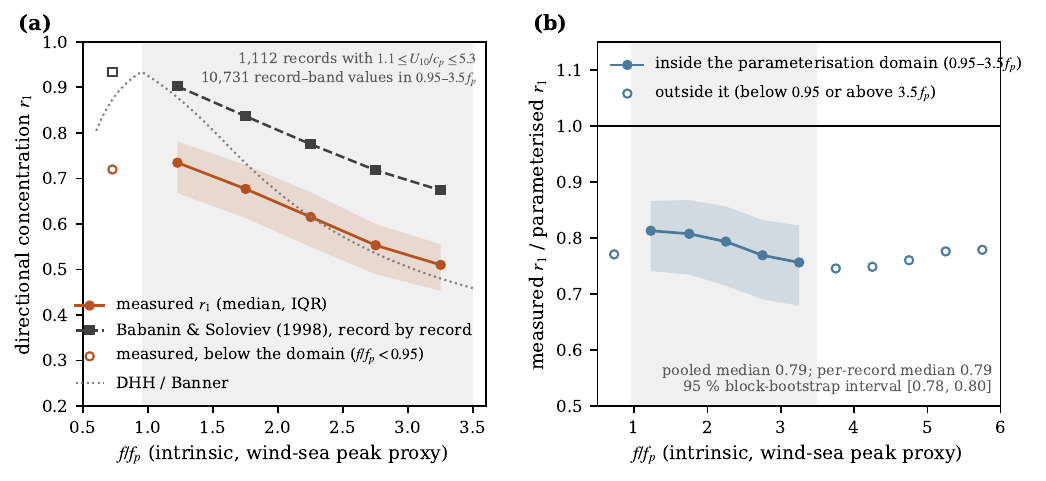}
  \caption{Measured directional concentration and the parameterisation of
  \citet{babanin1998directional}, for the $1{,}112$ records satisfying the wave-age criterion
  $1.1 \le \Uten/c_p \le 5.3$ at a wind-sea peak estimated within a restricted frequency window
  (selection in S3). (a) Median and interquartile range of the measured $r_1$ and the values
  calculated from the parameterisation in bins of $f/f_p$ (intrinsic frame); the first bin
  inside the parameterisation range spans $0.95$--$1.5 f_p$, the following bins are $0.5 f_p$
  wide. Dotted, the DHH form with the Banner extension. (b) Median and interquartile range of
  the ratio of measured to parameterised $r_1$, pooled over record--band values in each bin;
  filled symbols lie within the parameterisation's frequency range $0.95$--$3.5 f_p$, open
  symbols outside it. Results by wave-age class are given in S3.}
  \label{fig:r1_bs}
\end{figure}

Within the frequency and wave-age range of the wave-staff-array parameterisation of
\citet{babanin1998directional}, the measured directional concentration is lower but has a
similar dependence on relative frequency (Fig.~\ref{fig:r1_bs}). For each of the $1{,}112$
selected records we calculate the median ratio of measured to parameterised concentration
across the eligible frequency bands; the median of these ratios across records is $0.79$
($95\%$ block-bootstrap interval $0.78$--$0.80$). The comparison assumes a $\mathrm{sech}^2$
directional distribution when converting the published width to a concentration, and applies
only to the selected records; it does not independently validate the buoy moments.
Differences may reflect sea-state composition, the directional-reconstruction method
\citep{lin2022spreading,torres2025transfer} or measurement response; the tested additive-noise
model does not explain a difference of this size (S3).

\begin{figure}[tbp]
  \centering
  \includegraphics[width=\textwidth]{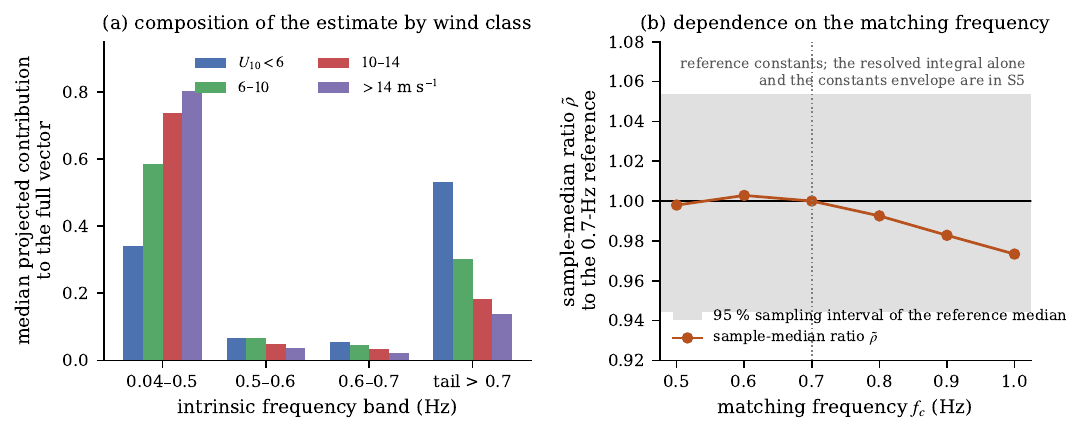}
  \caption{Measured and extrapolated contributions ($N = 13{,}139$). (a) Median fractional
  contribution along the direction of the total Stokes-drift vector, the projection of each
  band vector on that direction divided by the total magnitude, for the intrinsic bands $0.04$--$0.5$, $0.5$--$0.6$
  and $0.6$--$0.7$\,Hz and the tail above $0.7$\,Hz, by wind class. Medians of the contributions need
  not sum to one. (b) Sample-median ratio $\tilde\rho$ when the matching frequency is moved from the
  reference $0.7$\,Hz (dotted line), with the $95\%$ block-bootstrap sampling interval of the
  reference median (grey); the resolved integral alone and the
  scenario envelope of the tail constants are shown in S5.}
  \label{fig:tail}
\end{figure}

The reference estimate has a median speed $\hat u_s$ of $0.081$\,m\,s$^{-1}$ ($95\%$ interval
$0.077$--$0.086$) on the $13{,}139$ records, whose median wind speed is $6.8$\,m\,s$^{-1}$
(interquartile range $4.9$--$9.3$), and the median resolved speed $u_{\mathrm{res}}$ is
$0.052$\,m\,s$^{-1}$. The tail ratio $\Phi_{\mathrm{hf}}$ has a median of $0.37$
($0.33$--$0.41$) and decreases from $0.55$ in light winds to $0.14$ in strong winds (S6),
because $u_{\mathrm{res}}$ grows more strongly with wind speed than $u_{\mathrm{hf}}$.
Figure~\ref{fig:tail}a shows the projected band contributions by wind class; the frequencies below
$0.5$\,Hz account for about half of the projected contribution at the median. The median recordwise ratio of Stokes-drift speed to
wind speed is $0.0119$. The median absolute angle between the resolved Stokes vector and the
wind-sea direction is $15^\circ$, with an interquartile range of $7$--$26^\circ$. The comparison
with the wave models follows, first with gridded final vectors and then with matched
directional spectra, which separate the spectral level from the directional cancellation.

\subsection{Comparison with WAVEWATCH~III and MFWAM}
\label{sec:ww3}

\begin{figure}[tbp]
  \centering
  \includegraphics[width=\textwidth]{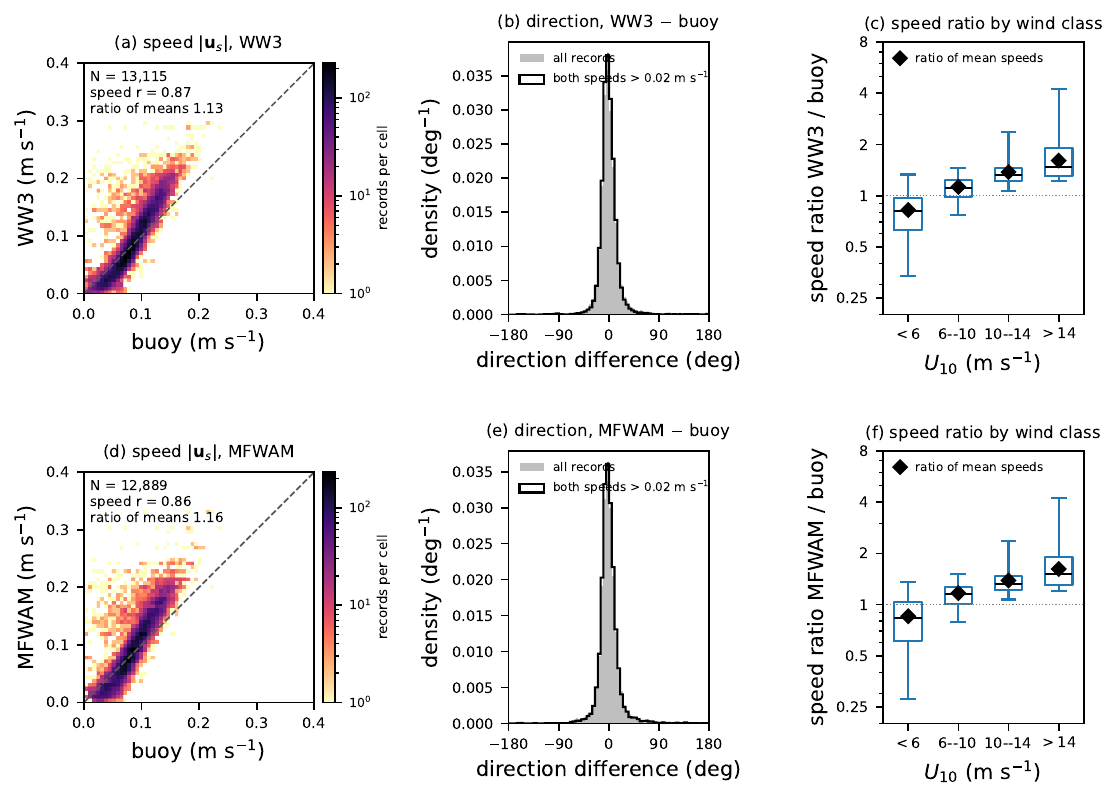}
  \caption{Modelled surface Stokes drift compared with the buoy-derived estimates: WAVEWATCH~III
  (a--c, $13{,}115$ records) and the Copernicus Marine MFWAM analysis (d--f, $12{,}889$ records).
  (a, d) Speed, observations on the horizontal axis, with the 1:1 line. (b, e) Wrapped difference
  of the propagation-toward compass bearings, model minus buoy, over the full $\pm180^\circ$
  range, for all records (mean absolute difference $14^\circ$ and $15^\circ$) and for the
  records with both speeds above $0.02$\,m\,s$^{-1}$ ($12{,}047$ and $11{,}913$ records;
  $11^\circ$ and $12^\circ$). (c, f) Model-to-buoy speed ratios by wind class (record counts
  $5{,}272$/$5{,}357$/$2{,}281$/$205$ for WW3 and $5{,}022$/$5{,}376$/$2{,}286$/$205$ for MFWAM):
  boxes span the interquartile range, whiskers the 5th--95th percentiles, diamonds the mean-speed ratios
  $\rho_{\mathrm{WW3}}$ and $\rho_{\mathrm{MF}}$ quoted in Sect.~\ref{sec:ww3}, whose block-bootstrap
  sampling intervals are given in S12. Component comparisons, storm-window time series and
  the full statistics tables are in S12.}
  \label{fig:ww3}
\end{figure}

The WAVEWATCH~III surface Stokes vectors covary strongly with the buoy-derived reference across
$13{,}115$ matched records (Fig.~\ref{fig:ww3}): the centred complex vector correlation is $0.92$, the
speed correlation $0.87$ and the mean absolute direction difference $14^\circ$ with a
negligible mean angular offset. Differences are expressed as model relative to buoy
throughout. The mean-speed ratio $\rho_{\mathrm{WW3}}$ is $1.13$ ($95\%$ interval $1.09$--$1.18$)
but varies with wind speed: it is $0.82$ in winds below $6$\,m\,s$^{-1}$ and
$1.13$, $1.38$ and $1.61$ in the $6$--$10$, $10$--$14$ and above-$14$\,m\,s$^{-1}$ classes.
The strongest-wind class contains $205$ records with a speed correlation of $0.30$.
Figure~\ref{fig:ww3}c shows $\rho_{\mathrm{WW3}}$ by wind class with the recordwise
distributions. Without the drift-Doppler correction ($D_{\mathrm{d}} = 1$) $\rho_{\mathrm{WW3}}$ would be about $1.19$, so the correction removes only part of the wind dependence.

Part of the remaining difference is one of definition. The buoy reference includes the
parameterised tail above $0.95$\,Hz, whose magnitude has a median ratio of $0.28$ to $\hat u_s$
($0.26$ of the mean), whereas the WW3 diagnostic ends at its last frequency bin, centred at
$0.95$\,Hz. Truncating the buoy estimate at $0.95$\,Hz, an approximately matched upper
frequency (the bin edges and integration weights of the two are not made identical), raises
$\rho_{\mathrm{WW3}}$ from $1.13$ to $1.49$ (to $1.47$ for a cut at $1$\,Hz):
within the range that the model resolves, WW3 exceeds the buoy estimate by about one half. The station-spectra comparison
below, on another sample and model configuration, likewise shows higher model estimates
within the compared band.

The Copernicus Marine MFWAM analysis shows a similar wind dependence (Fig.~\ref{fig:ww3}d--f).
Its mean-speed ratio $\rho_{\mathrm{MF}}$ is $1.16$, with a speed correlation of $0.86$; both
products give lower mean speeds than the buoy estimates in light winds and higher speeds in
stronger winds, and the pattern holds on the $12{,}865$ records where both products are
available (S12). The two products agree with each other far more closely than either does with
the buoy (MFWAM to WW3 ratio of mean speeds $1.02$, vector correlation $0.99$, direction
difference $7^\circ$), so the wind-dependent difference is not specific to one model product,
although the two share forcing information and unresolved-wave assumptions. Secondary statistics, and the bulk
relation of \citet{ardhuin2009observation} evaluated with the buoy's wind speed and wave
height for reference, are given in S12.

\begin{figure}[tbp]
  \centering
  \includegraphics[width=\textwidth]{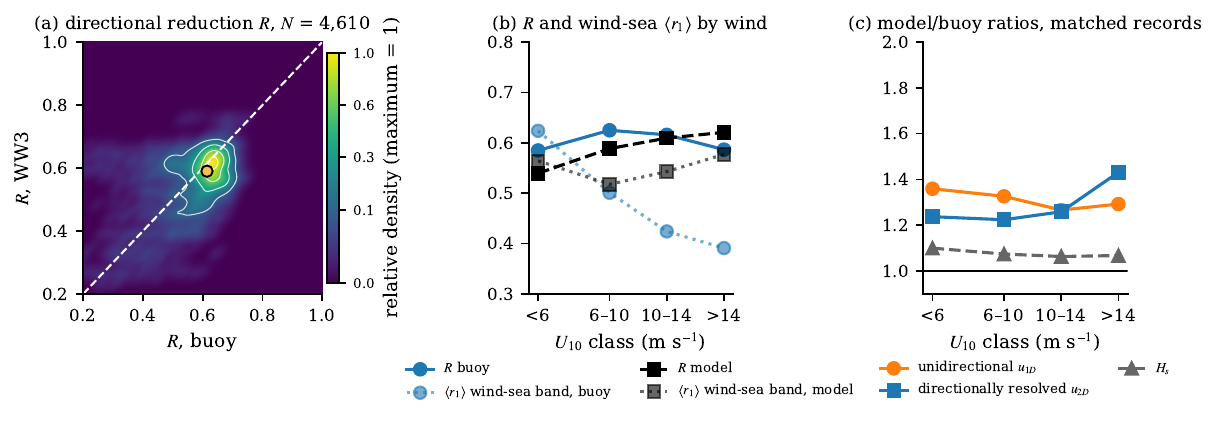}
  \caption{The model's directional spectra at the nearest hindcast station against the buoy
  records ($\le 60$\,km, $\le 90$\,min, $N = 4{,}610$, $4{,}383$ from OTC25-MELODI-04). (a)
  Directional reduction $R$ of the model spectra (vertical) against that of the buoy
  (horizontal), smoothed record density with contours at $10$, $30$, $60$ and $90\%$ of its
  maximum, the circle at the two medians. (b) $R$ and the Stokes-weighted concentration
  $\langle r_1\rangle$ over $0.60$--$0.90$\,Hz by wind class, buoy and model. (c) Median
  recordwise model-to-buoy ratios of the unidirectional-equivalent and directionally resolved
  Stokes estimates on comparable resolved bands (buoy $0.04$--$1$\,Hz, model bins to
  $0.95$\,Hz), and of the significant wave height.}
  \label{fig:ww3_stations}
\end{figure}

The gridded model fields deliver only the model's final vector, so a magnitude difference
cannot be attributed to the directional reduction or to the spectral energy level. The
model's directional spectra at its fixed output stations make this separation possible. For
each of the $4{,}610$ matched pairs of buoy and model spectra ($30$ stations, $4{,}383$ records
from OTC25-MELODI-04 and none from OTC25-MELODI-20; Fig.~\ref{fig:ww3_stations}) the Stokes
drift is calculated with and without its directional information. Including wave directions
leaves a median fraction $R = 0.61$ of the unidirectional-equivalent estimate for the buoy
spectra and $0.59$ for the model spectra ($R_{\mathrm{sp}}$ $0.66$ against $0.64$, $R_{\mathrm{md}}$
$0.94$ against $0.93$); the buoy and model reductions agree within $0.03$ overall and within
$0.05$ in every wind class. Different directional reductions therefore explain little of the
difference in estimated speed on this predominantly Biscay sample. Equal reductions do not
prove equal directional distributions.

Despite their similar directional reductions, the buoy and model spectra give different
Stokes-drift magnitudes. Over the analysis band the model's $u_{\mathrm{1D}}$ and $u_{\mathrm{2D}}$ are $1.31$ and $1.24$
times the buoy's (medians of recordwise model-to-buoy ratios), and the median ratio of model to buoy significant wave
height is $1.08$, approximately independent of wind speed in this sample. These results point to lower spectral energy
levels in the buoy records at the frequencies that carry the Stokes drift, most of all above
$0.7$\,Hz, rather than to a large difference in directional cancellation.
Below $0.7$\,Hz, which carries a fraction $0.91$ of the buoy's drift resolved to $1$\,Hz, the
model-to-buoy Stokes ratio of $1.19$ is consistent with the variance ratio of $1.16$ implied by
the wave-height ratio. The discrepancy is larger above $0.7$\,Hz, where the buoy carries $0.09$ of its
resolved drift and the model $0.14$; the band-by-band ratios are given in S12. Pilot,
Spotter and sampling-emulation comparisons suggest a contribution from the sampling and
filtering response of the recovered buoy's $3.2$\,Hz channel, but do not uniquely
distinguish buoy-response errors from model errors (S11). None of these comparisons validates the extrapolated tail independently; its sensitivity to
the assumed spectrum and directionality follows.

\subsection{Sensitivity to the unresolved short waves and to the spreading function}
\label{sec:sens}

The estimate is more sensitive to the assumed directional spreading of the tail than to modest
changes in the matching frequency (Table~\ref{tab:unc}). Moving the matching frequency from
$0.7$ to $0.6$ or $0.8$\,Hz gives $\tilde\rho = 1.003$ and $0.993$ (Fig.~\ref{fig:tail}b). This stability reflects the calibration of the tail constants over
$0.6$--$0.7$\,Hz; it does not establish the tail's accuracy. If the directional distribution
continues to broaden above the matching frequency as in \citet{banner1990equilibrium}, scaled
to the measured concentration at $f_c$, $\rho = 0.91$, and $0.81$ in light winds. In the isotropic-tail scenario, waves above the wind-dependent
wavenumber of \citet{lenain2020hfstokes} are distributed equally in all directions, so their
contributions give no net Stokes drift; for winds above about $10$\,m\,s$^{-1}$ the whole
parameterised tail lies in this range, and $\rho = 0.78$ (S6).
Directions change by less than $3^\circ$ in either case. The spectral level of the tail
matters as well: the published constants of \citet{toba1973local} and
\citet{phillips1985equilibrium} give $\tilde\rho = 0.86$, and the drift-Doppler
assumptions give $\rho$ within $0.99$--$1.01$ (Table~\ref{tab:unc}, S4--S5). These tests
quantify the sensitivity to specified alternatives, not confidence bounds on the true surface
Stokes drift, and their effects are not combined.

\begin{table}[t]
\centering\footnotesize
\caption{Sampling uncertainty and sensitivities of the reference estimate ($N = 13{,}139$,
median magnitude $0.081$\,m\,s$^{-1}$). Sampling entries are $95\%$ block-bootstrap intervals
of the sample median. Other entries are the sample-median ratio $\tilde\rho$ or the median recordwise ratio $\rho$ of
Eq.~\eqref{eq:ratios}. They are not accuracy estimates. Breaking-induced drift is a physical contribution outside the
estimated quantity (S10).}
\label{tab:unc}
\begin{tabular}{L{0.13\textwidth}L{0.40\textwidth}L{0.17\textwidth}L{0.18\textwidth}}
\toprule
kind & source & ratio & value \\
\midrule
sampling & block bootstrap of the median, constants refitted & $\tilde\rho$ & $0.94$--$1.06$ \\
 & joint bootstrap interval of $(\beta_T, \alpha_P)$ & $\tilde\rho$ & $0.987$--$1.012$ \\
matching frequency & $0.6$ / $0.8$\,Hz instead of $0.7$; $1.0$\,Hz & $\tilde\rho$ & $1.003$ / $0.993$; $0.974$ \\
tail level & published constants $(0.06, 8.3\times10^{-3})$ & $\tilde\rho$ & $0.86$ \\
 & calibration band $0.6$--$1.0$\,Hz, adjusted for the channel steepening & $\tilde\rho$ & $1.033$ \\
tail spreading & continued Banner broadening & $\rho$ & $0.91$ ($0.81$ in light winds) \\
 & isotropic above $f_M = 0.42/\ustar$ & $\rho$ & $0.78$ \\
drift Doppler & relative buoy speed $\times 0.7$ / $\times 1.3$; advecting current; directional reconstruction & $\rho$ & $0.989$ / $1.012$; $0.990$; $0.998$ \\
excluded & breaking-induced drift (S10) & -- & not estimated \\
\bottomrule
\end{tabular}
\end{table}

\begin{figure}[tbp]
  \centering
  \includegraphics[width=\textwidth]{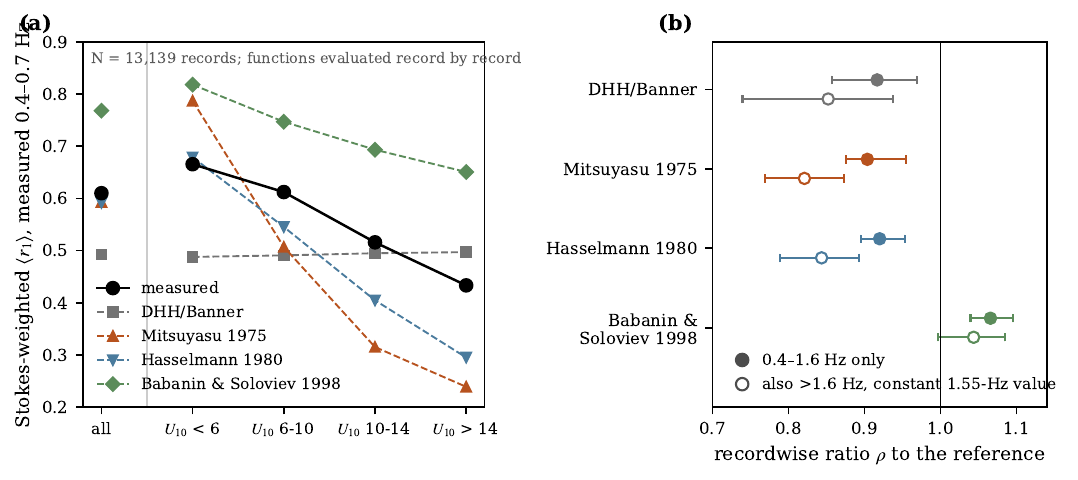}
  \caption{Spreading parameterisations against the measured concentration ($N = 13{,}139$).
  (a) $\langle r_1\rangle_{0.4\text{--}0.7}$ by wind class, measured
  and from four parameterisations evaluated record by record; the value for all records is
  shown apart from the connected wind classes. (b) Median recordwise ratio $\rho$ when the reference concentration is replaced over $0.4$--$1.6$\,Hz only
  (filled) or when, in addition, the concentration above $1.6$\,Hz is replaced by the function's
  value at $1.55$\,Hz, held constant (open); bars show the interquartile range of the recordwise
  ratios. The reference concentration is measured below $0.7$\,Hz and parameterised above it.}
  \label{fig:spreading}
\end{figure}

Short waves make a substantial contribution to the unidirectional-equivalent surface
estimate, and only part of it is observed. Waves between $0.4$ and $1.6$\,Hz, wavelengths of
$10$\,m to $0.6$\,m, account for about two fifths of the estimate (Fig.~\ref{fig:weight}b):
the median fractional contributions across records are $0.19$ from the measured
$0.4$--$0.7$\,Hz band and $0.21$ from the parameterised $0.7$--$1.6$\,Hz part of the tail,
with $0.40$ below $0.4$\,Hz and $0.18$ above $1.6$\,Hz (separately computed medians, which need not sum to
one). The contribution from $0.4$ to $0.7$\,Hz is obtained from the measured spectrum, where
the buoy constrains the directionality; the higher-frequency contribution is parameterised. The measured
$\langle r_1\rangle_{0.4\text{--}0.7}$ decreases with wind speed, from $0.67$ below $6$\,m\,s$^{-1}$ to $0.43$ above $14$\,m\,s$^{-1}$ (Fig.~\ref{fig:spreading}a).
In this sample the inverse wave age $\Uten/c_p$ is strongly correlated with the wave-derived wind speed, and
the two dependences are not separated (S6).

We evaluated four published spreading parameterisations record by record with the same
spectral weights and wave conditions: the $\mathrm{sech}^2$ form of DHH with the Banner
extension, the wave-age-dependent $\cos^{2s}$ forms of \citet{mitsuyasu1975directional} and
\citet{hasselmann1980directional}, and the integral width of \citet{babanin1998directional}
(S6). Over the measured $0.4$--$0.7$\,Hz band the DHH form has no explicit wind or wave-age
parameter and gives a nearly constant concentration across these wind classes, the two
$\cos^{2s}$ forms predict a steeper decrease with wind speed than observed for the sea-state
inputs and extrapolations used here, and the Babanin and Soloviev form is narrower than the
observations throughout (Fig.~\ref{fig:spreading}a); the wind-sea peak proxy sits at the edge
of its search window for $60\%$ of the records, and several forms are evaluated outside their
original domains (S6). Replacing the reference concentration, that is the measured concentration below $0.7$\,Hz
and the parameterised tail concentration above it, by each function produces both decreases
and increases of the estimated speed. The median recordwise ratio $\rho$ for the four functions ranges from $0.90$ to $1.07$ when only
$0.4$--$1.6$\,Hz is replaced, and from $0.82$ to $1.04$ when, in addition, the reference concentration of the tail above $1.6$\,Hz is
replaced by the function's value at $1.55$\,Hz, held constant at higher frequencies
(Fig.~\ref{fig:spreading}b; values per function in S6). The second experiment changes the
concentration of the remaining tail; it does not extrapolate the frequency dependence of the
function to infinity. These ranges span the median ratios of the four functions, not record-to-record scatter, and
because contributions from different frequency ranges compensate, the smallest speed change
does not identify the most accurate directional model.

\section{Discussion}
\label{sec:discussion}

The matched directional spectra help distinguish differences in spectral energy from
differences in directional cancellation. On this predominantly Biscay sample the median
retained fractions are similar for the buoy and model spectra, $0.61$ and $0.59$, so their
different Stokes-drift magnitudes appear to arise mainly from spectral energy differences
within the matched band. The comparisons suggest a contribution from the measurement
response at high frequencies, but do not uniquely assign the discrepancy to buoy or model
error. The gridded comparison also depends on how each product treats the unresolved tail:
the distributed WW3 diagnostic omits an analytical tail above its last bin at $0.95$\,Hz,
MFWAM adds an $f^{-5}$ tail above $0.58$\,Hz, and the buoy-derived estimate adds a calibrated
equilibrium-to-saturation tail above $0.7$\,Hz. The pooled $\rho_{\mathrm{WW3}}$ of
$1.13$ for the full buoy estimate thus combines a lower buoy level within the resolved range
with a tail contribution that the diagnostic omits, and it averages opposite-signed
differences across wind classes.

The wind-class differences describe this sample and do not define a correction to either the
buoy or the model estimates. In light winds the tail dominates the
buoy estimate and $\rho_{\mathrm{WW3}} = 0.82$ is of the order of the
directional-tail scenarios, so the light-wind excess of the buoy may be a property of the tail
rather than of the resolved measurement. In strong winds the buoy estimate is below the model in the resolved band
itself, on a small sample with low speed correlation. Because wind speed and friction
velocity are inferred from the wave records and also enter the estimator through the wind
classes, the relative buoy velocity and the tail level, the wind-class differences are descriptive model comparisons and not an independent wind-response relationship. The limited strongest-wind
sample and the predominantly Biscay matched spectra restrict their generality further.

Both effects, directional cancellation and the weight of the short-wave tail, are known; the
contribution here is their observational characterisation with compact drifting wave
buoys and the matched model-spectrum comparison, which provides information absent from a
comparison of final Stokes vectors alone. The retained fraction of $0.61$ is consistent with
earlier observations for the conditions and frequency range analysed here: \citet{webb2015spreading}
obtained $0.62$ from the Ocean Weather Station~P buoy, and the present hindcast spectra give
$0.59$. What the frequency-resolved moments add is a constraint on directional
concentration between $1$ and $10 f_p$, where above about $2 f_p$ the observed median lies
between the DHH plateau and the Banner extension and near the peak below both. The tail
follows the equilibrium-to-saturation form of \citet{lenain2020hfstokes}, calibrated on the
measured band (S5). The telemetry
archive also illustrates a separate band-averaged approximation where directional moments are
unavailable; its configuration-specific calibration is discussed in S8.

The observations constrain part of the short-wave spectrum along the drifter trajectories
and help diagnose differences in model-derived Stokes estimates. Their value lies in the
measured frequency-dependent moments, not in complete resolution of the shortest waves: the
unresolved contribution remains dependent on a spectral continuation and its assumed
directionality. The scales that carry the largest part of that continuation are also those at which spectral wave models depend most on parameterised source terms and on the
resolution of the forcing winds, and at which in situ observations are scarcest. The comparison with the bulk relation of
\citet{ardhuin2009observation}, driven by the buoy's wind speed and wave height, illustrates
the information lost when the observed spectral structure is replaced by aggregate wave
parameters, although the shared inputs prevent an independent validation (S12).

Three limitations bear on the full surface estimate: the tail calibration uses the buoy's own
measured band, the noise tests are conditional and do not bound every field error of the
moments at $0.6$--$0.9$\,Hz, and the buoy velocity relative to the water of $0.019\,\Uten$ is
a model assumption. They do not remove the observational value of the resolved
spectra, but they prevent a unique attribution of the model--observation differences. For the
conversion of a unidirectional-equivalent estimate to a vector, the decrease of the measured
concentration between $0.4$ and $0.7$\,Hz with wind speed is the property that a spreading
function must reproduce, and none of the four tested forms does so with the sea-state inputs
and extrapolations used here. Estimates that account for the modulation of the short waves by the orbital
velocities of the long waves, set aside here, are left to a separate study.

Finally, the estimated surface Stokes drift is not, by itself, the Stokes contribution to the
buoy's motion. Averaging over the hull draft is a kinematic approximation, and the available
response information does not identify a steady transport coefficient (S9). Stokes drift and
direct wind forcing also have similar directions and wind dependence, so their contributions
are difficult to distinguish from these trajectories alone, as documented in the leeway
literature \citep{sutherland2020leeway}. Separating them requires independent observations or
additional constraints.

\section{Conclusions}
\label{sec:conclusions}

Compact drifting wave buoys provide frequency-dependent directional information for
estimating the surface Stokes drift. Across $13{,}139$ records the median estimated
Stokes-drift speed is $0.081$\,m\,s$^{-1}$ at a median wind speed of $6.8$\,m\,s$^{-1}$.
Accounting for wave directions reduces the magnitude by a median $39\%$ over the measured
spectral band of $0.04$--$1$\,Hz. The median ratio of the parameterised tail magnitude above $0.7$\,Hz to the total magnitude is $0.37$,
and assumptions about the tail's directional spreading matter more than
modest changes in the matching frequency.

WAVEWATCH~III and MFWAM speeds are strongly correlated with the buoy-derived estimates, with
similar wind-dependent differences for the two products. On the matched, predominantly Biscay
sample, buoy and model spectra show similar directional reductions, so spectral energy
differences account for much of the discrepancy within the compared band, although
buoy-response and model errors cannot be uniquely separated.

These new observations provide constraints for evaluating modelled surface Stokes drift in a
spectral range weakly constrained by observations, without establishing the accuracy of the
complete surface estimate. The unresolved short-wave contribution remains parameterised. Better knowledge of the short waves and of the buoy's high-frequency response would strengthen the full-spectrum estimates; the separate Stokes contribution to the buoy's own
motion remains unidentified.

\section*{CRediT authorship contribution statement}
\textbf{Alexey S. Mironov:} Conceptualization, Methodology, Software, Formal analysis,
Investigation, Data curation, Writing -- original draft, Visualization.
\textbf{Fabrice Collard:} Investigation, Writing -- review \& editing.
\textbf{Gw\'ena\"ele Jan:} Data curation, Writing -- review \& editing.
\textbf{Bertrand Chapron:} Supervision, Writing -- review \& editing.

\section*{Declaration of competing interest}
A.~Mironov and G.~Jan are employed by eOdyn, which manufactures the MELODI drifter. F.~Collard is
employed by OceanDataLab. B.~Chapron declares no competing interest.

\section*{Data availability}
\begingroup\sloppy
The quality-controlled trajectories of the OTC25 MELODI fleet, including the three deployments
analysed here, with positions, sea-surface temperature, drift, significant wave height, wave
period and the buoy-derived wind, are published on SEANOE \citep[doi:10.17882/\allowbreak 117337]{mironov2026otc25data}.
The per-record surface Stokes drift reference estimates of this paper for the $13{,}139$ records
(surface vector, resolved vector to $0.7$\,Hz, tail ratio $\Phi_{\mathrm{hf}}$ and direction, drift-Doppler
factor $D_{\mathrm{d}}$, directional reductions, quality flags), with the run
configuration (wind-product revision, frequency conventions, matching frequency, tail constants,
moment definitions), with the collocated WAVEWATCH~III and MFWAM fields, are deposited as a companion dataset on the same portal (CF NetCDF and
CSV) \citep[doi:10.17882/\allowbreak 118145]{mironov2026stokesdata}. The raw $3.2$\,Hz motion records, the directional
moments and the analysis code are available from the corresponding author on request.
\endgroup

\section*{Supplementary material}
Supplementary sections S1--S12 (derivations, sensor and data-tier details, validation of the
directional moments, drift-Doppler and tail-calibration tests, closure scenarios, exploratory
analyses and the full comparison statistics) are appended to this preprint after the reference
list, with their own page numbers, figure and table numbering (S) and bibliography; in the journal
version they form a separate file.

\section*{Acknowledgements}
The authors are grateful to Lucas Charron (eOdyn) for the design, development, manufacturing and
testing of the MELODI buoys, and to the wider eOdyn team for their help with the MELODI project and
for their administrative and technical support. They thank Craig Donlon (ESA) for his help and
support during the ESA Ocean Training Course 2025 (OTC25) deployment campaign. The WAVEWATCH~III hindcasts GLOBMULTI\_\allowbreak ERA5\_\allowbreak GLOBCUR\_01 (fields) and
GLOB-30M\_\allowbreak LOPS\_2025 (station spectra) were produced by Ifremer/LOPS and are distributed through the Ifremer Sextant
catalogue. The MFWAM wave analysis was provided by the E.U. Copernicus Marine Service and ERA5
by the Copernicus Climate Change Service.
Funding: the OTC25 buoys were funded by the Centre National d'\'Etudes Spatiales, CNES [contract
5700012680]; this work was partially supported by the AI4COPSEC project under the European Union
Horizon Europe programme [grant 101190021]. The funders had no role in the study design, the
analysis or the decision to submit.

\section*{Declaration of generative AI and AI-assisted technologies in the writing process}
During the preparation of this work the authors used Claude (Anthropic) to help edit the
manuscript text and to draft the implementation of the analysis code and the plotting code. After
using this tool, the authors reviewed and edited the content as needed and take full responsibility
for the content of the published article.

\bibliographystyle{elsarticle-harv}
\bibliography{references}

\clearpage


\section*{Supplementary material}

# Supplementary material to "Surface Stokes drift from compact drifting wave buoys"

Alexey S. Mironov[a], Fabrice Collard[b], Gwénaële Jan[a], Bertrand Chapron[c]

[a] *eOdyn, Technopôle Brest-Iroise, 115 rue Claude Chappe, 29280 Plouzané, France*
[b] *OceanDataLab, 870 route de Déolen, 29280 Locmaria-Plouzané, France*
[c] *Ifremer, Univ. Brest, CNRS, IRD, Laboratoire d'Océanographie Physique et Spatiale (LOPS), ZI de la Pointe du Diable, CS 10070, 29280 Plouzané, France*

**Abstract**

This supplement collects the derivations, validation details and exploratory analyses that support the main text: the noise weighting of the acceleration form (S1), the sensor payload (S2), the directional-moment checks (S3), the tests of the drift-Doppler correction (S4), the calibration of the parameterised high-frequency tail and its deployment-holdout tests (S5), the directional closure scenarios (S6), the analytic parametric spectrum and the DHH normalisation (S7), the parameterised spreading used for the telemetered dataset (S8), the kinematic draft factor (S9), nonlinear and higher-order effects (S10), the open-ocean telemetry buoys (S11) and a glossary of the model-to-buoy ratios (S12). Sections S1–S6 are the methods behind the reference estimate; S8, S11 and S12 hold additional observations and comparisons; S7, S9 and S10 are exploratory physical context that is not needed to follow the main results. Terms follow the main text: the reference estimate is the measured spectrum to an intrinsic frequency of 0.7 Hz plus the parameterised tail; the analysis band is the whole measured band 0.04–1 Hz; the unidirectional-equivalent magnitude is the estimate from the omnidirectional spectrum with all waves in one direction; directional cancellation is the loss of magnitude when the measured directions are taken into account, and $R$ is the fraction of the unidirectional-equivalent magnitude retained, so that the lost fraction is $1 - R$. The resolved and tail speeds $u_{\mathrm{res}}$ and $u_{\mathrm{hf}}$, the tail-to-total magnitude ratio $\Phi_{\mathrm{hf}}$ (tail ratio), the drift-Doppler factor $D_{\mathrm{d}}$, and the ratios to the reference estimate $\tilde{\rho}$ (sample median), $\rho$ (recordwise

*Email address:* alexey.mironov@eodyn.com (Alexey S. Mironov)

median) and $\rho_{\rm m}$ (model to buoy mean speeds) are defined in Sect. 2.2 of the main text (Eqs. 2 and 3). Section, equation and figure numbers of the form "Sect. 2.2" or "Eq. (2)" refer to the main text.

---

## S1. The acceleration form and the weighting of sensor noise

Equation (1) of the main text is the Kenyon integral written in terms of the vertical-acceleration spectrum $S_{\rm acc}(f) = (2\pi f)^4 E(f)$. It follows from the monochromatic result: for a deep-water wave of amplitude $a$, wavenumber $k$ and angular frequency $\omega = 2\pi f$ the surface Stokes drift is $\omega k a^2$ along the wave vector, components add to second order in slope, and with $a^2/2 \to E\, df\, d\theta$ and $k = (2\pi f)^2/g$ the prefactor $2\omega k$ becomes $16\pi^3 f^3/g$; since $16\pi^3/(2\pi)^4 = 1/\pi$ the same integral reads $(1/\pi g) \int S_{\rm acc}(f) f^{-1} [a_1 + {\rm i} b_1]\, df$. The surface value is set by the third spectral moment and the Stokes transport $\int \boldsymbol{u}_s\, dz = 2\pi \int f E [a_1 + {\rm i} b_1]\, df$ by the first, so the surface drift depends on much higher frequencies than the transport, the wave height or the energy (Breivik et al., 2014). For a record from a single accelerometer it is the same integral as the displacement form, term by term, including its noise (Fig. S1b): the transformation is algebraic and does not change the measurement. Its use is that the frequency weighting of an additive noise contribution becomes explicit. Let the acceleration record contain an additive white noise floor of spectral level $N_a$ that does not alter the normalised moments. Its contribution to the vector between $f_{\min}$ and a cutoff $f_c$ is

$$
\begin{aligned}
\delta \boldsymbol{u}_s &= \frac{N_a}{\pi g} \int_{f_{\min}}^{f_c} \frac{a_1(f) + {\rm i}\, b_1(f)}{f}\, df, \\
|\delta \boldsymbol{u}_s| &\le \frac{N_a}{\pi g} \int_{f_{\min}}^{f_c} \frac{r_1(f)}{f}\, df \;\le\; \frac{N_a}{\pi g} \ln \frac{f_c}{f_{\min}},
\end{aligned}
\tag{S1}
$$

so that the bound grows logarithmically with the cutoff and the equality with a "band-mean $r_1$" holds only for aligned directions and the logarithmic weighting. A white noise floor of level $N_v$ in a velocity record enters as $E = N_v/(2\pi f)^2$, an integrand $4\pi f N_v/g$ that grows with frequency and gives an error quadratic in $f_c$ (Fig. S1c). The two noise levels have different physical units, so Fig. S1c compares the shape of the two kernels and not the performance of two sensors. When noise also changes the normalised moments, lowering $r_1$ without a compensating term in the autospectra, (S1)

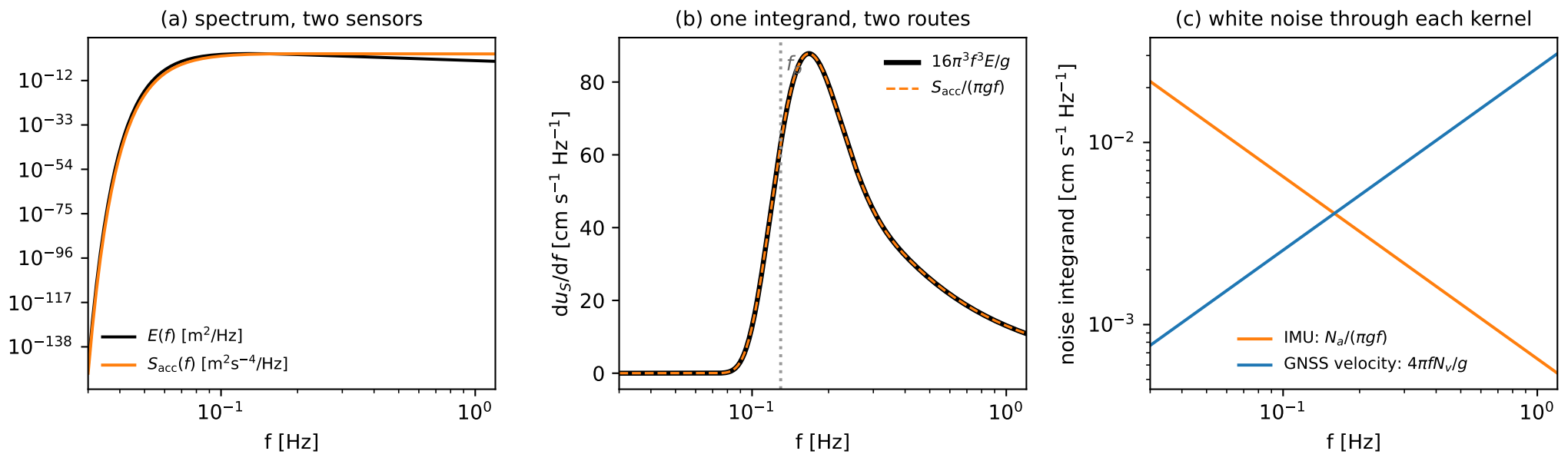


Figure S1: One integrand, two forms. (a) Displacement and acceleration spectra of a Donelan–Hamilton–Hui wind sea at $U_{10} = 10\,\mathrm{m\,s^{-1}}$. (b) The Kenyon integrand in displacement form, $16\pi^3 f^3 E/g$, and in acceleration form, $S_{\mathrm{acc}}/(\pi g f)$, identical by construction; the Stokes-drift contribution is concentrated above the spectral peak $f_p$. (c) A white noise floor of unit level in each record, mapped through its kernel: acceleration noise enters as $1/f$, velocity noise as $f$. The two levels have different units; the panel compares kernel shapes.

is not the complete bias; that part is assessed under tested conditions in S3, which does not supply a general bound on field errors.

*Finite depth.* The estimator uses deep-water dispersion. For the DHH wind sea at $U_{10} = 6$–$18\,\mathrm{m\,s^{-1}}$, with and without a swell partition ($T_p = 12$–$16\,\mathrm{s}$, 20–50% of the variance), the finite-depth surface integral, with $\omega^2 = gk \tanh kh$ and the kernel $\cosh(2kh)/(2\sinh^2 kh)$, exceeds the deep-water one by a factor of at most 1.01 at $h = 100\,\mathrm{m}$, 1.04 at 50 m and 1.12 at 30 m. These are idealised unidirectional spectra, not a bound for the observed directional records, and the track depths were not screened.

## S2. Platform, sensors, data tiers and wind product

The MELODI hull is a disk of two half-shells, 240 mm in diameter and 100 mm in height, moulded from the biodegradable polyester polybutylene succinate, with a 200 mm stabilising pole and ballast (Fig. S2). The payload comprises a multi-constellation GNSS receiver (u-blox M10S), a nine-axis inertial measurement unit, a sea-surface temperature sensor, a solar array charging lithium-ion cells and an Iridium Short Burst Data modem (Table S2). Acquisition runs in records of about 22 min every 30 min. On board, an attitude filter reconstructs the Earth-frame vertical acceleration and the microcontroller computes its spectrum over 0.02–1 Hz in 128 bands and the

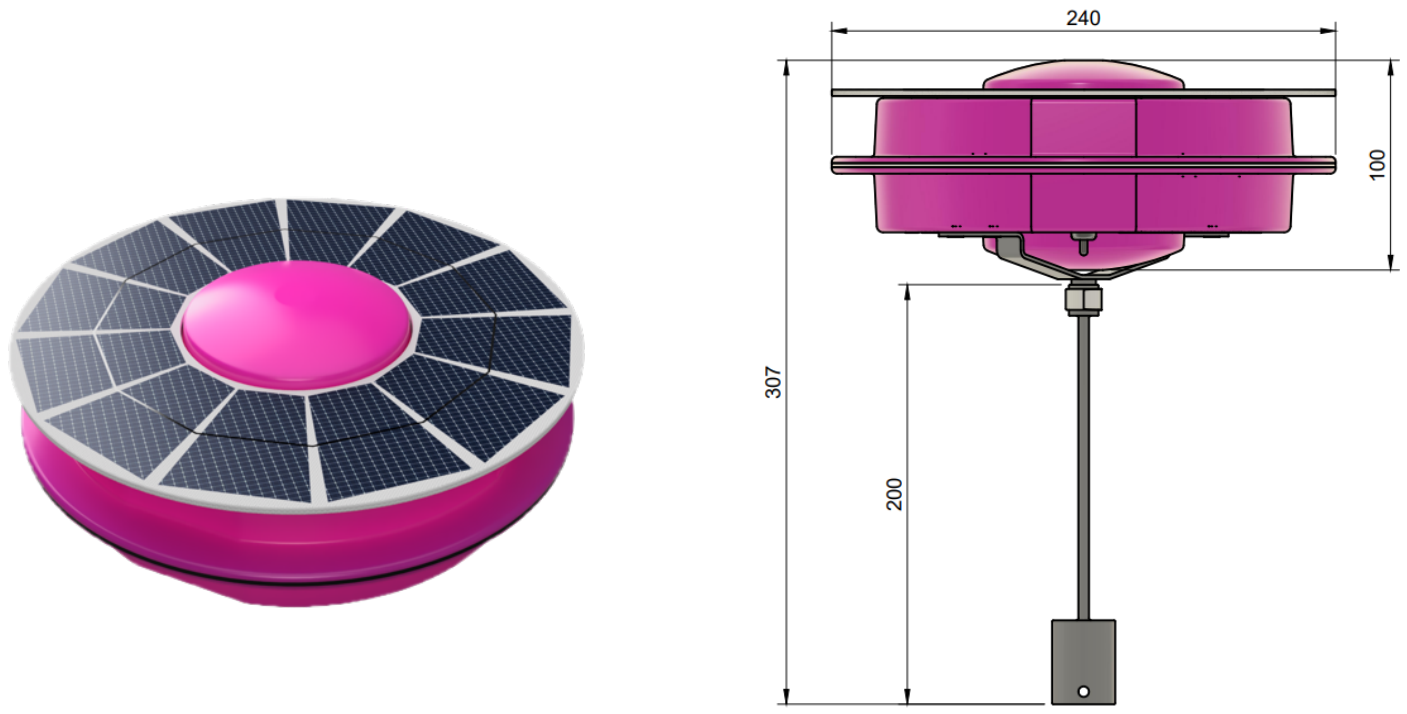


Figure S2: The MELODI drifter: exterior view with the solar array (left) and cross-section with principal dimensions in millimetres (right).

significant wave height $H_s = 4\sqrt{\int E\,df}$, transmitted by Iridium at a cadence of one or three hours together with 10-min positions. On the recovered fleet the memory card holds the filtered attitude and the raw inertial channels at 3.2 Hz; the pilot hardware logs the inertial unit at 100 Hz and GNSS Doppler velocities at 4 Hz. Between March 2025 and March 2026, 28 buoys were deployed from the Norwegian Sea to the equatorial Indian Ocean, and four have been recovered with their full records.

*Data tiers.* The observations come in four tiers (Table S1). The telemetry tier (T1) archives an omnidirectional acceleration spectrum, so it gives a unidirectional-equivalent estimate, and a vector only with a parameterised spreading (S8). The recovered tier (T2) adds the directional moments; per-frequency moments are available on 13,139 of its 21,742 records, from three of the four recovered deployments, and band-averaged moments on all of them; the fourth recovered deployment, COI_2025 (1.5°N, Somali basin, 181 days), enters only the band-averaged archive. The pilot tier (T3) supplies the 100 Hz accelerometer used to check the spectral slope of the recovered buoys' 3.2 Hz channel (S11). The fourth tier (T4), two drifters running the firmware of August 2026, deployed west of Biscay, transmits onboard displacement spectra and first moments below 0.5 Hz (S11). The distribution of the distances of the records matched to the hindcast stations is shown in Fig. S3.

*Wind product and friction velocity.* Wind speed is taken from the MELODI wind product, revision 1 of 27 August 2026: a ridge regression with a gated

Table S1: Data tiers. Records are 22-minute acquisitions; T4 records are 30-minute cycles.

| tier | recorded | size | supports |
|---|---|---|---|
| T1 telemetry | archived omnidirectional $S_{\mathrm{acc}}(f)$, 0.006–1.025 Hz ($\Delta f \approx 6.3$ mHz); buoy-derived wind speed, direction, $u_*$ | 8,245 spectra, 24 buoys, 2025–2026 | unidirectional-equivalent estimate; vector with parameterised spreading and a quality flag |
| T2 recovered | filtered attitude and raw inertial channels at 3.2 Hz, GNSS positions; $a_1(f), b_1(f)$ and $S_{\mathrm{acc}}(f)$ recomputed | 4 deployments, 21,742 records; 13,139 with per-frequency moments on 3 deployments | the vector estimate, tail calibration, model comparison |
| T3 pilot | 4 Hz GNSS Doppler velocities and 100 Hz inertial unit (deployment 090326, buoys N1–N4, Iroise Sea) | 48 h, 91 records after quality control, $H_s$ to 3.6 m | spectral slope to 1 Hz |
| T4 open-ocean telemetry | onboard displacement spectra and first moments to 0.5 Hz, buoy wind (two new-firmware drifters west of Biscay, August 2026) | 3 days, 125 records after quality control, $H_s \approx 1$ m | directional moments below 0.5 Hz in the open ocean |

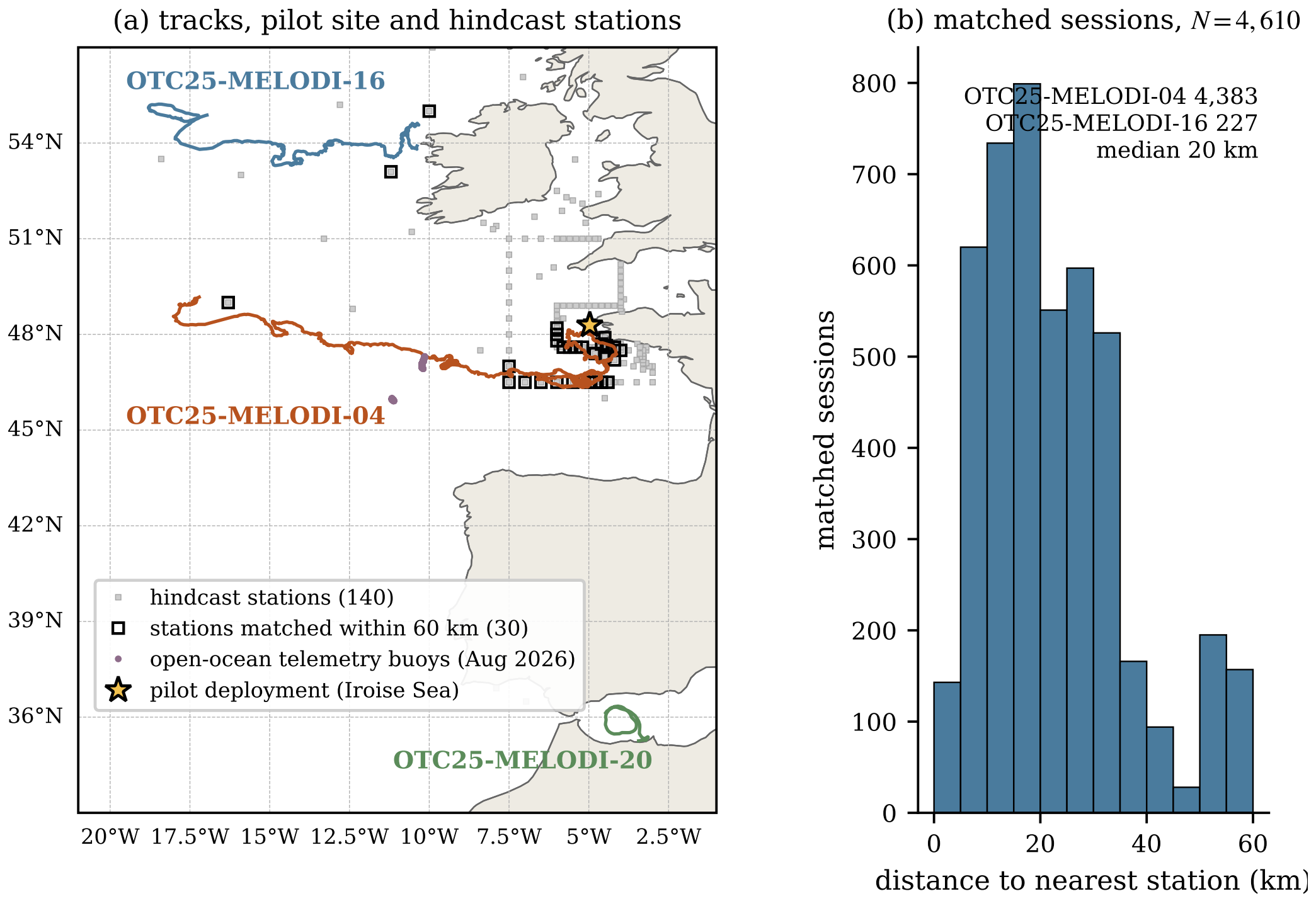


Figure S3: (a) Tracks, pilot site and hindcast stations as in Fig. 1 of the main text. (b) Distance of the 4,610 matched records to the nearest station.

Table S2: MELODI sensor payload. Inertial channels run at 3.2 Hz on the recovered fleet (T2) and 100 Hz on the pilot hardware (T3). The GNSS receiver gives 10-min positions on the fleet and 4 Hz Doppler velocities on the pilot and new-firmware hardware.

| sensor | range | resolution | rate |
|---|---|---|---|
| GNSS receiver | – | 2.5 m | 0.1–4 Hz |
| accelerometer | ±4 g | $0.061 \times 10^{-3}$ g | 3.2–100 Hz |
| gyroscope | $\pm 250^\circ\,s^{-1}$ | $4.4 \times 10^{-3}\,^\circ\,s^{-1}$ | 3.2–100 Hz |
| magnetometer | $\pm 800\,\mu T$ | $0.02\,\mu T$ | 3.2 Hz |
| sea-surface temperature | −10 to 85 °C | 0.06 °C | 0.03 Hz |

gradient-boosting residual correction on nine acceleration-spectrum features, calibrated on scatterometer collocations and validated leave-one-buoy-out to a root-mean-square difference of $0.90\,m\,s^{-1}$ (Mironov, 2026). The wind-sea direction used for the validation of that product is the despiked and smoothed peak direction of the 0.60–0.90 Hz band from the gyroscope-derived slopes, retrieved to a mean absolute error of 7–13°; the tail moment of the estimator is the acceleration-weighted complex band mean over the same band (main text, Eq. 2). The friction velocity is the linear Toba inversion of the 0.275–0.475 Hz acceleration level with a coefficient of 0.062. An earlier retrieval version stored with the archived products differs from revision 1; its effect on the tail fit is given in S5.

## S3. Directional-moment checks and measurement limitations

*Spectral processing.* The per-record spectra and moments of the reference estimate were computed from the 3.2 Hz records of the recovered buoys as follows (the settings are those of the code that produced the companion dataset). Inputs are the attitude-filtered vertical acceleration (heave channel), the raw gyroscope pitch and roll rates and the attitude heading; records shorter than 410 samples are skipped at reading, and the estimator requires at least one full Welch segment of 819 samples (about 4.3 min), which all retained records satisfy. The rates are integrated to slope angles with a centred cumulative sum (see the timing paragraph below) and high-passed with a fourth-order zero-phase Butterworth filter at 0.02 Hz; the slopes are rotated with the heading into east and north surface slopes, and the acceleration has its record mean removed. The three channels then pass through one estimator: samples deviating from the channel mean by more than ten standard

deviations are replaced by the mean; a first-order recursive high-pass filter with a 4 s time constant (corner near 0.04 Hz) is applied; Welch segments of 256 s (819 samples) advanced by 204 samples (75% overlap) are linearly detrended, multiplied by a sine taper and rescaled to their pre-taper variance; the auto- and cross-spectra are averaged over the segments, merged over three adjacent bins (spacing $3/256 \approx 0.0117$ Hz) and retained to 1.1 Hz. With $S_{xx}, S_{yy}, S_{zz}$ the east-slope, north-slope and acceleration autospectra and $S_{xz}, S_{yz}$ the slope–acceleration cross-spectra, the first moments are the quadrature ratios

$$a_1 = \frac{\operatorname{Im} S_{xz}}{\sqrt{(S_{xx} + S_{yy})\, S_{zz}}}, \qquad b_1 = \frac{\operatorname{Im} S_{yz}}{\sqrt{(S_{xx} + S_{yy})\, S_{zz}}}, \tag{S2}$$

each clipped to $[-1, 1]$; the sign convention gives $a_1 + ib_1$ in the propagation-toward direction, verified against the wind direction in the wind-sea band. Bands with a non-positive acceleration or slope autospectrum are excluded, and a record is retained when at least five bands remain in 0.04–1 Hz and a wind-product speed is available. Record accounting: the recovered archive holds 21,742 records from four deployments; per-frequency moments exist for 13,139 records from three deployments (S2); 13,115 have a collocated WW3 field, 12,889 an MFWAM field and 4,610 a station spectrum (main Sect. 2.1); the tail calibration uses the records whose projected 0.6–0.7 Hz contribution exceeds $10^{-4}\,\mathrm{m\,s^{-1}}$ (S5); the S6 closure scenarios exclude two records with zero retrieved wind speed, whose peak-frequency proxy requires $U_{10} > 0$, and these records remain in the released product and in the all-record summaries. The companion dataset carries three quality flags, which do not exclude records from the statistics of this paper: a degraded wind retrieval, a concentration $r_1 < 0.2$ in 0.60–0.90 Hz (swell–wind-sea transitions, tail direction unreliable) and a missing friction velocity or tail moment.

*Electronic floor.* Static on-deck records of the attitude unit (a sample of 2,542 archived records of OTC25-MELODI-04) give a white electronic floor of $4.5 \times 10^{-7}$ $(\mathrm{m\,s^{-2}})^2\,\mathrm{Hz^{-1}}$ in acceleration and $1.8 \times 10^{-8}$ $(\mathrm{rad\,s^{-1}})^2\,\mathrm{Hz^{-1}}$ in the gyroscope. These are static records rather than a documented deck calibration, and the quality control retains records whose acceleration variance is below the wave-motion threshold. At sea the 5%-quantile spectrum ($H_s \approx 0.85$ m) falls as $f^{-3.3}$ in acceleration and $f^{-3.5}$ in the gyroscope and lies $10^5$ to $10^6$ above the electronic floor at 0.6–0.9 Hz.

*Synthetic recovery.* Three spreading cases (DHH/Banner, the fleet's measured $r_1(f)$, and a broad case) at six wave heights from 0.3 to 4.5 m were synthesised with exact slope rates, integrated with the centred scheme, high-passed and passed through the Welch cross-spectral estimator, 200 realisations each, with the static floor added. The median ratio of recovered to prescribed $r_1$ is 1.02 in 0.3–0.5, 0.5–0.7 and 0.7–0.9 Hz, and the 5–95% range of single realisations is 0.95–1.11, 0.94–1.13 and 0.93–1.15 in the three bands. The excess ratio of 1.01–1.02 is consistent with the finite-record bias of the magnitude of a mean unit vector, which is present without noise; the widest-spread fleet case at $H_s = 4.5$ m reaches a median of 1.05 in 0.7–0.9 Hz. $R_{\mathrm{sp}}$ is recovered at 1.01 and $R$ at 1.00 (0.94–1.04). A paired comparison of noiseless and noisy realisations with identical wave fields, which isolates the noise contribution from the finite-record dispersion, has not been run; the numbers above describe the sum of both under the tested conditions. For independent additive noise in the slope and acceleration channels the population cross-spectrum is unchanged while both autospectral factors of the denominator increase; with $s_s$ and $s_a$ the signal-to-noise power ratios of the summed slope channels and of the acceleration, the attenuation is

$$\frac{r_{1,\mathrm{obs}}}{r_{1,\mathrm{true}}} = \left[(1+s_s^{-1})(1+s_a^{-1})\right]^{-1/2}, \tag{S3}$$

a population expectation to which finite-record dispersion and the magnitude bias add. The coloured-noise simulations add noise to both channel groups. Equal ratios $s_s = s_a = s$ give $s/(1+s)$ and noise in the slope channels alone $\sqrt{s_s/(1+s_s)}$, so a reading of $r_1$ lower by a factor 0.65–0.75 at 0.6–0.9 Hz would require $s = 1.9$–$3.0$ in the equal-ratio case or $s_s = 0.7$–$1.3$ for slope-only noise, orders of magnitude below the ratios implied by the static electronic floor. That floor cannot explain the observed departure; the comparison does not bound motion-dependent errors or differential channel response.

*Quadrature check.* The coherence-type ratio $\gamma$, with $\gamma^2 = (|S_{xz}|^2 + |S_{yz}|^2)/[(S_{xx} + S_{yy})S_{zz}]$, is formed from the same cross-spectra and the same denominator as $r_1$. After the centred integration the median $r_1/\gamma$ is 0.96 over 0.60–0.90 Hz; by wind class the pair $r_1/\gamma$ is 0.64/0.66, 0.53/0.55, 0.44/0.47 and 0.40/0.44. Because $\gamma$ shares numerator and denominator with $r_1$, additive slope noise lowers both alike, and the check tests phase consistency, not the presence of noise.

*Band-averaged moments.* Band-averaged moments interpolated to the grid give magnitudes closely correlated with those from the per-frequency moments ($r = 0.97$) but a biased direction, so per-frequency moments are used wherever direction matters.

*Comparison with the Babanin–Soloviev parameterisation.* Fig. 4 of the main text compares the measured concentration with Babanin and Soloviev (1998), whose Eq. (19) gives the integral width $A^{-1} = \int K\, d\theta$ of the directional distribution normalised to unit maximum as $A = 1.12\,(U_{10}/c_p)^{-0.50}(f/f_p)^{-0.95} + (2\pi)^{-1}$ for $f \geq 0.95 f_p$, from Black Sea wave-staff arrays analysed with a maximum-likelihood method, for $1.1 \leq U_{10}/c_p \leq 5.3$ and $0.7 \leq f/f_p \leq 3.5$. $A$ is converted to $r_1$ through the sech$^2$ shape, $A = \beta_s/[2\tanh(\pi\beta_s)]$; an integral width alone does not fix the first moment of an arbitrary distribution, so the conversion is an assumption. The comparison is frame-consistent: the wind-sea peak proxy, the elevation-spectrum maximum in 0.15–0.6 Hz, is transformed to the intrinsic frame with the Doppler shift of its own bin, and $f/f_p$, $U_{10}/c_p = 2\pi f_p U_{10}/g$ and the 0.05-Hz bands are all intrinsic. Only the 5,275 records whose proxy lies away from the lower window edge are used; a maximum inside the window is a proxy for the wind-sea peak, not an established partition. Inside the parameterisation domain (1,112 records: 845 OTC25-MELODI-04, 251 OTC25-MELODI-16, 16 OTC25-MELODI-20; 10,731 record–band values, 5% above 0.7 Hz) the pooled record–band median of the ratio of measured to parameterised $r_1$ is 0.79 (quartiles 0.71, 0.85) and the median of the per-record medians 0.79 (deployment-stratified 3-day-block bootstrap, 300 replicates, 95% interval 0.78–0.80; leave-one-deployment-out 0.79–0.80); at or below 0.7 Hz the pooled ratio is 0.79, above it 0.79. With the parameterisation extrapolated, the pooled ratio is 0.75 over 3.5–6$f_p$ and 0.79 for $U_{10}/c_p < 1.1$ (4,163 records); the class breakdown is shown in Fig. S4, where the sparsely populated highest class (99 records) spans 0.67–0.80. Under the independent additive-noise model of Eq. (S3), a concentration ratio of 0.79 corresponds to $s \approx 3.8$ with equal ratios ($s_s \approx 1.7$ for slope-only noise), against $10^5$–$10^6$ implied by the static floor at sea; the test is conditional on that noise model.

*Directional concentration against frequency, full diagnostic.* Figure S5 repeats Fig. 3 of the main text with the coherence-type ratio $\gamma$ and the high-peak-frequency subset ($f_p \geq 0.15$ Hz, 1,377 records), which was examined for the sensitivity to low-frequency spectral peaks and decreases faster, from

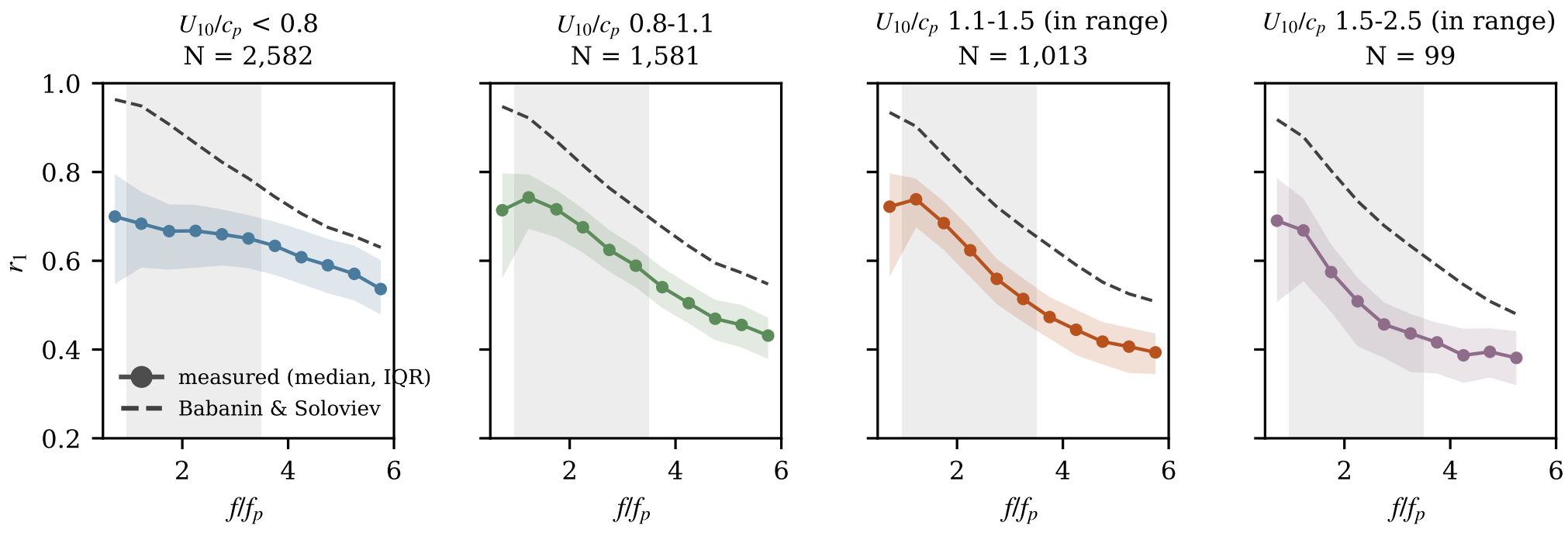


Figure S4: Measured $r_1$ (median and interquartile range) against $f/f_p$ by inverse-wave-age class at the intrinsic wind-sea peak proxy, records with the proxy resolved, with the Babanin–Soloviev parameterisation evaluated record by record (dashed); the grey band is its validity range in $f/f_p$, and only the two right-hand classes lie inside its wave-age range.

0.71 at 1.6–2$f_p$ to 0.47 at 5–6.5$f_p$. A timing error would have biased $r_1$ low: a right-Riemann cumulative sum of the gyroscope rates lags the acceleration channel by half a sample and rotates the slope–heave cross-spectrum out of quadrature by $\pi f\,\Delta t$, a factor $\cos(\pi f\,\Delta t)$ on $r_1$, 0.71 at 0.8 Hz, which the centred integration removes.

## S4. Tests of the drift-Doppler correction

The drift-Doppler correction of Sect. 2.2 assigns one Doppler factor $(f_i/f_e)^3$ to the mean direction of each measured band, with the slip $0.019\,U_{10}$ along the record's Stokes direction and $U_{10}$ from the wind product. Its size on the resolved vector to intrinsic 0.7 Hz, the drift-Doppler factor $D_{\rm d}$ of Sect. 2.2, is 1.09 at the median, 1.06, 1.10, 1.12 and 1.14 in the four wind classes, and 1.10 on the analysis band to 1 Hz. Scenario vectors were recomputed from the same cross-spectra on the intrinsic support of the reference estimator, combined with the reference tail held fixed, and summarised by the recordwise ratio $\rho$ with the 5–95% range of the recordwise ratios: $\rho = 0.989$ (0.973–1.000) for the slip ×0.7, 1.012 (1.000–1.029) for the slip ×1.3, 0.999 (0.986–1.001) for the slip direction taken from the wind sea instead of the Stokes vector, and 0.990 (0.977–1.001) for a wind-driven surface current of $0.005\,U_{10}$ advecting the waves. The slip scenarios grow with wind, from ratios of $1 \mp 0.005$ in calm to $1 \mp 0.03$ in strong winds.

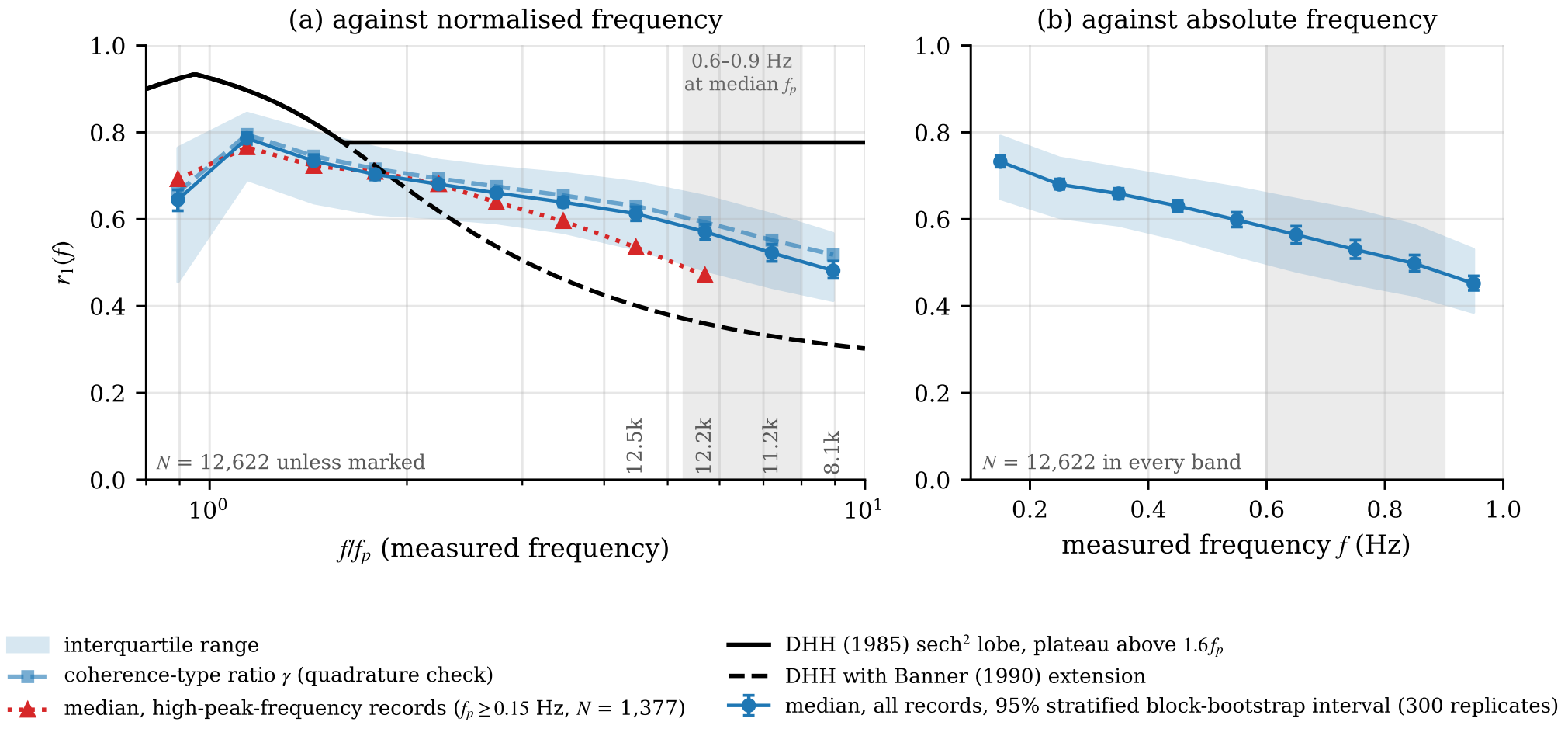


Figure S5: Directional concentration $r_1$ against $f/f_p$ (a) and absolute frequency (b), as Fig. 3 of the main text, with the coherence-type ratio $\gamma$ formed from the same slope–heave cross-spectra and the high-peak-frequency subset ($f_p \geq 0.15\,\mathrm{Hz}$); bins with fewer records than the full sample are marked.

The direction-dependent transformation was tested with two reconstructions of the directional distribution: a cos-2s distribution from the first moments on all records and a maximum-entropy distribution from the first and second moments on a wind-stratified subsample of 500 records. Evaluated as complete vectors on the intrinsic support of the reference estimator with the tail held fixed, $\rho = 0.998$ (5–95% range 0.991–1.008) for cos-2s and 0.998 (0.988–1.035; the cos-2s reconstruction on the same 500 records gives 0.998) for maximum entropy, and the resolved-vector direction changes by 0.3°; the unidirectional-equivalent estimate changes by a factor 1.03–1.04, because spreading dilutes the Doppler factor in the scalar but not in the moments.

The frequency support was also tested. Selecting the resolved bands by encounter frequency $f_e \leq f_c$ includes waves above intrinsic 0.7 Hz while the analytic tail starts at intrinsic 0.7 Hz. Selecting them by intrinsic frequency $f_i \leq f_c$, as the reference estimator does, gives resolved vectors to 0.7 and 1.0 Hz of 0.974 and 0.984 of their encounter-selected values, uniformly across wind classes, and the tail constants were fitted on the intrinsic 0.6–0.7 Hz band. The same intrinsic edges are used for the per-band comparison with the station spectra of Sect. 3.2.

## S5. Tail calibration and deployment-holdout tests

The closed-form tail is $\Delta u_s^{\mathrm{hf}}(f_c) = 2\beta_T u_* \ln(f_t/f_c) + \alpha_P g/(\pi f_t)$ with $f_t = \alpha_P g/(2\pi\beta_T u_*)$ from continuity of the two ranges; if $f_t \le f_c$ the first term is dropped and $f_c$ replaces $f_t$ in the second. For reference, Lenain and Pizzo (2020) give the fraction of the unidirectional surface Stokes drift above a cutoff $f_c$ as $1.3315\, e^{-2.47 f_c}$, that is 0.24 at 0.7 Hz, and define their constants on the omnidirectional wavenumber spectrum, a saturation range $Bk^{-3}$ with $B = 7 \times 10^{-3}$ and an equilibrium range $(\beta/2)(u_*/\sqrt{g})k^{-5/2}$ with $\beta = 0.105$; in the frequency form used here these correspond to $\alpha_P = 2B = 0.014$ and $\beta_T = 0.105$, both about 1.7 times the published values of Phillips (1985) and Toba (1973). Their separately fitted transition, $k_n u_*^2/g = 0.0097$, corresponds to $f_n \approx 0.154/u_*$ Hz ($u_*$ in $\mathrm{m\,s^{-1}}$ in all such expressions), whereas continuity of the two ranges with these constants would place it at $0.21/u_*$ Hz. With the fitted constants the transition is $f_t \approx 0.30/u_*$ Hz, above the matching frequency for winds below about $11\,\mathrm{m\,s^{-1}}$, so the equilibrium segment contributes in calm and moderate winds and the saturation segment alone in strong winds.

The constants $(\beta_T, \alpha_P)$ are fitted by directional closure on the 0.6–0.7 Hz intrinsic band. With $\boldsymbol{t} = e^{\mathrm{i}\theta^{\mathrm{ws}}}$ the unit vector in the tail direction of the record, the target is the measured band contribution projected onto that direction and the prediction is the unidirectional tail integral over the band multiplied by the tail concentration,

$$
\begin{aligned}
q_{\mathrm{obs}} &= \mathrm{Re}\left\{\left[\hat{\boldsymbol{u}}_s^{\mathrm{res}}(0.7) - \hat{\boldsymbol{u}}_s^{\mathrm{res}}(0.6)\right]\boldsymbol{t}^*\right\}, \\
q_{\mathrm{pred}} &= r_1^{\mathrm{ws}} \int_{0.6}^{0.7} \frac{16\pi^3}{g} f^3 E_{\mathrm{tail}}(f; u_*, \beta_T, \alpha_P)\, df,
\end{aligned}
\qquad \text{(S4)}
$$

where $\hat{\boldsymbol{u}}_s^{\mathrm{res}}(f)$ is the drift-Doppler-corrected resolved vector to intrinsic frequency $f$. The constants minimise the sum over four wind classes of the squared logarithm of the median ratio $q_{\mathrm{pred}}/q_{\mathrm{obs}}$, using the records with $q_{\mathrm{obs}} > 10^{-4}\,\mathrm{m\,s^{-1}}$ and the classes with more than 20 such records, on a grid of 0.005 in $\beta_T$ and 0.0005 in $\alpha_P$ refined locally to 0.001 and 0.0001. The fitted pair therefore describes closure under the prescribed tail direction and concentration and absorbs the measured spectral level and the processing response; it is not an independent estimate of universal equilibrium-range constants. The wind classes use the wind product of Sect. 2.1 (revision 1); the archived speeds of an earlier retrieval version differ from it by a median absolute $0.52\,\mathrm{m\,s^{-1}}$ and move 14% of the records across a class boundary,

and with them the fit would give $(0.069, 1.42 \times 10^{-2})$ and $\tilde{\rho} = 1.026$, so the wind-product revision is part of the reference configuration. The reference fit is $(0.069, 1.32 \times 10^{-2})$, with closure ratios of 0.97, 1.05, 0.94 and 1.03 in the four wind classes; the fitted $\alpha_P$ is 1.6 times the standard value and 0.94 of the value implied by Lenain and Pizzo (2020); "published constants" below denotes the standard pair $(0.060, 8.3 \times 10^{-3})$. A stratified block bootstrap of 200 refits, resampling 3-day blocks within each deployment, gives 0.068–0.071 and 1.27–1.37 $\times 10^{-2}$; the joint propagation of the resampled pairs to the sample median magnitude gives $\tilde{\rho} = 0.987$–1.012. Fitted on two deployments and applied to the third, the constants are $(0.069, 1.26 \times 10^{-2})$ for OTC25-MELODI-04 held out, $(0.070, 1.34 \times 10^{-2})$ for OTC25-MELODI-16 and $(0.069, 1.32 \times 10^{-2})$ for OTC25-MELODI-20, so one held-out constant falls just below the bootstrap interval; the held-out closure ratio is 0.98, 1.05 and 0.86, and the held-out median magnitude is $\tilde{\rho} = 0.984$, 1.006 and 1.000 of the all-data value. These holdout changes compare two calibrations on the same held-out records and test the sensitivity to the calibration sample; small changes in the total estimate do not imply exact agreement in the calibration band, as the Alboran Sea deployment shows with no change in median speed but a held-out closure ratio of 0.86. Fitting on all but one calendar month gives constants within (0.069–0.070, 1.31–1.33 $\times 10^{-2}$) and held-out closure ratios of 0.96–1.03. The Alboran Sea record, where calm winds dominate and the equilibrium segment contributes most, is the least well described.

Two alternative closure bands were tested. Fitting on 0.6–1.0 Hz with the 0.7–1.0 Hz part of the buoy spectrum corrected for the steepening of the recovered buoy's 3.2 Hz channel (S11) by the difference between the pilot-tier slope (−4.34) and the recovered-tier slope (−5.57) gives $(0.067, 1.46 \times 10^{-2})$ and $\tilde{\rho} = 1.033$; its agreement is conditional on that correction. Fitting on 0.6–1.0 Hz as measured gives $(0.059, 1.27 \times 10^{-2})$ and $\tilde{\rho} = 0.980$, the steepened band pulling the saturation level down. With the published constants $(0.060, 8.3 \times 10^{-3})$ the median is $0.070\,\mathrm{m\,s^{-1}}$, $\tilde{\rho} = 0.86$, and $\rho_{\mathrm{WW3}} = 1.27$. Across the range of published constants $\beta_T \in [0.04, 0.08]$, $\alpha_P \in [6, 11] \times 10^{-3}$ $\tilde{\rho}$ lies within 0.80–0.94. Figure S6 shows the median speed against the matching frequency with the resolved integral alone and the scenario envelope $\beta_T \in [0.04, 0.08]$, $\alpha_P \in [0.006, 0.016]$, which is a sensitivity envelope and not a sampling interval.

Multiplying $\alpha_P$ by 1.1 gives $\tilde{\rho} = 1.035$, and multiplying $\beta_T$ by 1.1 gives 1.002; the median per-record fractional sensitivities (relative change of the estimate per relative change of the constant) are 0.29 and 0.01.

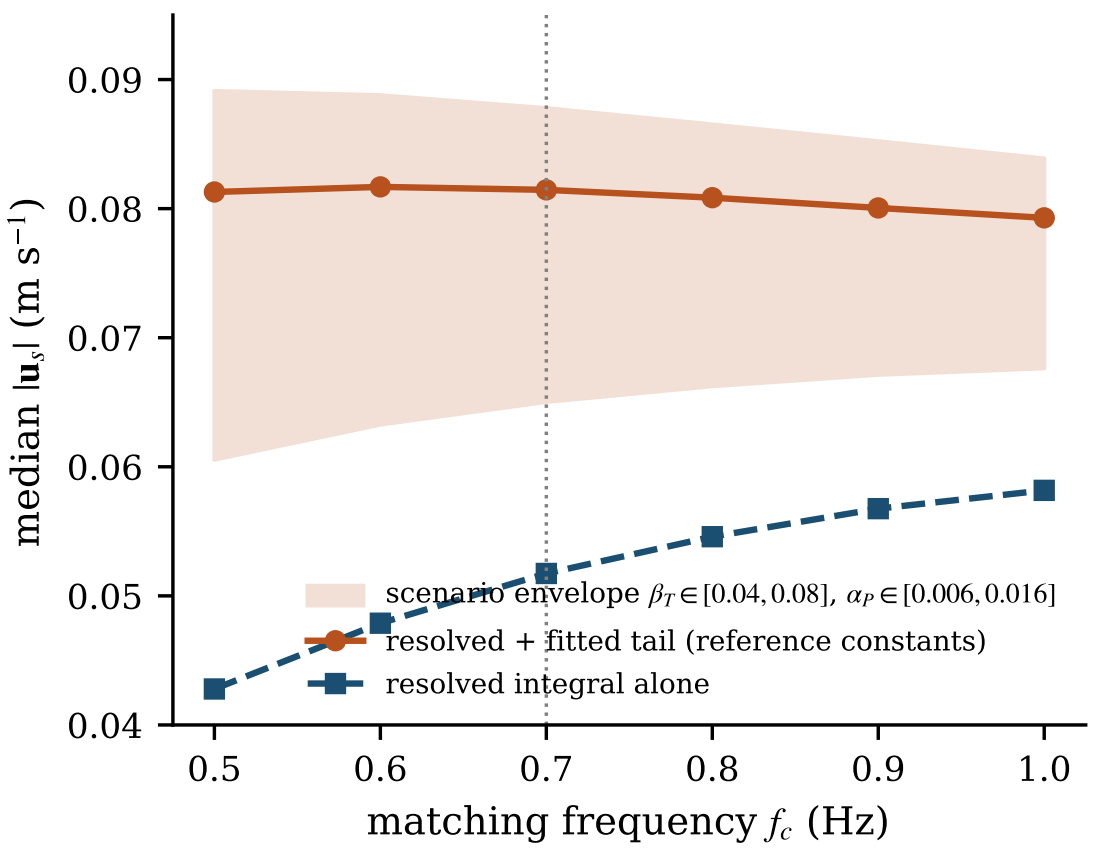


Figure S6: Median speed of the estimate against the matching frequency: resolved integral plus the fitted tail (reference constants), the resolved integral alone, and the scenario envelope of the tail constants. The dotted line marks the reference 0.7 Hz.

## S6. Directional closure scenarios of the tail

The reference tail extends to infinity with the direction $\theta^{\text{ws}}$ and concentration $r_1^{\text{ws}}$ of Sect. 2.2, the argument and magnitude of the acceleration-spectrum-weighted complex first moment over the measured 0.60–0.90 Hz band. Two alternatives were evaluated for every record with the reference constants and matching frequency, integrating the parametric tail analytically to infinity for the constant-direction and isotropic cases and numerically on a logarithmic grid to $10^4$ Hz for the Banner case (Table S3). Scenario (ii) continues the Banner (1990) broadening: the $\text{sech}^2(\beta_s\theta)$ lobe with $\beta_s(x) = 10^{-0.4+0.8393\exp[-0.567\ln(x^2)]}$, $x = f/f_p > 1.6$, is normalised on the circle $[-\pi, \pi]$ and its first moment $r_1$ computed numerically, the same convention as in Fig. 3 of the main text (the unbounded-angle form $u/\sinh u$ with $u = \pi/(2\beta_s)$ gives 0.22 instead of 0.30 at $10f_p$); $f_p$ follows the fully developed relation $f_p = 0.14\,g/U_{10}$, the curve is scaled to the measured $r_1$ at $f_c$, and its limiting value is 0.25 on the circle. Scenario (iii) takes the field as isotropic above the wavenumber $k_M = (g/u_*^2)\exp[(\pi/2-\theta_0)/\gamma]$ of Lenain and Pizzo (2020), with $\theta_0 = 2.835$ and $\gamma = 0.48$, that is above $f_M = 0.418/u_*$ Hz ($u_*$ in m s$^{-1}$), and constant $r_1$ below.

Under scenario (iii) the tail contributes no vector where $f_M \leq f_c$, which is the case for the median record above 10 m s$^{-1}$. The two scenarios are not confidence limits: they are medians of recordwise changes under the paper's

Table S3: Directional closure scenarios of the tail, reference configuration ($f_c = 0.7$ Hz, $\beta_T = 0.069$, $\alpha_P = 1.32 \times 10^{-2}$), medians over records: the tail ratio $\Phi_{\mathrm{hf}}$ (tail above $f_c$, to infinity), the median recordwise ratio $\rho$ under scenarios (ii) and (iii), the absolute direction change, and $f_M$. $\Phi_{\mathrm{hf}}$ is a ratio of magnitudes; Fig. 5 of the main text shows projected contributions.

| wind class | $N$ | $\Phi_{\mathrm{hf}}$ | $\rho$ (ii) | $\rho$ (iii) | $\Delta\theta$ (ii) | $\Delta\theta$ (iii) | $f_M$ (Hz) |
|---|---|---|---|---|---|---|---|
| all | 13,137 | 0.37 | 0.91 | 0.78 | 0.8° | 2.5° | 1.03 |
| $U_{10} < 6$ | 5,270 | 0.55 | 0.81 | 0.77 | 2.1° | 2.5° | 1.87 |
| 6–10 | 5,376 | 0.31 | 0.93 | 0.77 | 0.7° | 2.8° | 0.89 |
| 10–14 | 2,286 | 0.19 | 0.97 | 0.82 | 0.3° | 2.0° | 0.66 |
| $> 14$ | 205 | 0.14 | 0.98 | 0.86 | 0.2° | 1.6° | 0.60 |

own tail forms and describe the size of what the buoy does not measure. The scenarios exclude two records with zero retrieved wind speed, whose peak-frequency proxy requires positive wind speed.

*Spreading functions for the* 10 *m to* 0.6 *m waves.* For Sect. 3.3 and Fig. 8 of the main text, four published spreading parameterisations are evaluated record by record on the intrinsic 0.05-Hz bands, with the intrinsic wind-sea peak proxy $f_p$ and $U_{10}/c_p = 2\pi f_p U_{10}/g$ (quartiles 0.55, 0.77, 1.03): the DHH $\mathrm{sech}^2$ lobe with the Banner extension as above; the $\cos^{2s}$ form of Mitsuyasu et al. (1975) with $s = s_p(f/f_p)^{-2.5}$ above the peak and $s_p = 11.5\,(U_{10}/c_p)^{-2.5}$; the $\cos^{2s}$ form of Hasselmann et al. (1980) with $s = 9.77\,(f/f_p)^{\mu}$ above the peak and $\mu = -2.33 - 1.45\,(U_{10}/c_p - 1.17)$; and the integral width of Babanin and Soloviev (1998) converted to $r_1$ as in S3. For $\cos^{2s}$, $r_1 = s/(s+1)$, and $s$ is clipped to $[0.3, 50]$, so that $r_1$ lies in $[0.23, 0.98]$; this floor is an implementation choice where the forms are extrapolated beyond their observed range, and the extrapolated forms cannot approach isotropy. The baseline is the reference estimate (resolved vector to intrinsic 0.7 Hz, parametric tail from 0.7 Hz with the reference constants and the reference tail direction and concentration). Two experiments replace the concentration: (i) within 0.4–1.6 Hz only, that is in the measured 0.4–0.7 Hz bands, keeping their measured directions, and in the 0.7–1.6 Hz part of the parametric tail evaluated at 0.1-Hz nodes, with the tail above 1.6 Hz unchanged; (ii) in addition above 1.6 Hz, where the reference tail concentration is replaced by the function's value at the last node, 1.55 Hz, held constant to infinity; the frequency dependence of the function is not continued beyond 1.6 Hz. The statistic is the median recordwise ratio $\rho$ (Tables S4 and S5; interquartile ranges in Fig. 8b). The

Table S4: Stokes-weighted directional concentration over the measured 0.4–0.7 Hz band, measured and from four spreading parameterisations evaluated record by record, by wind class ($N$ = 5,272 / 5,376 / 2,286 / 205); B&S = Babanin and Soloviev (1998).

| $U_{10}$ (m s$^{-1}$) | measured | DHH/Banner | Mitsuyasu | Hasselmann | B&S |
|---|---|---|---|---|---|
| all | 0.61 | 0.49 | 0.59 | 0.59 | 0.77 |
| $< 6$ | 0.67 | 0.49 | 0.79 | 0.68 | 0.82 |
| 6–10 | 0.61 | 0.49 | 0.51 | 0.55 | 0.75 |
| 10–14 | 0.52 | 0.49 | 0.32 | 0.40 | 0.69 |
| $> 14$ | 0.43 | 0.50 | 0.24 | 0.30 | 0.65 |

Table S5: Median recordwise ratio $\rho$ when the reference concentration is replaced over 0.4–1.6 Hz only / with the tail concentration above 1.6 Hz also replaced by the constant 1.55-Hz value, by wind class.

| $U_{10}$ (m s$^{-1}$) | DHH/Banner | Mitsuyasu | Hasselmann | B&S |
|---|---|---|---|---|
| all | 0.92 / 0.85 | 0.90 / 0.82 | 0.92 / 0.84 | 1.07 / 1.04 |
| $< 6$ | 0.85 / 0.72 | 0.96 / 0.79 | 0.94 / 0.80 | 1.05 / 1.00 |
| 6–10 | 0.93 / 0.88 | 0.88 / 0.81 | 0.91 / 0.84 | 1.07 / 1.05 |
| 10–14 | 0.99 / 0.97 | 0.90 / 0.86 | 0.92 / 0.89 | 1.08 / 1.07 |
| $> 14$ | 1.02 / 1.01 | 0.92 / 0.90 | 0.93 / 0.91 | 1.08 / 1.07 |

sample-median ratio $\tilde{\rho}$ is within 0.01 of $\rho$ for every form. Inverse wave age at the wind-sea peak proxy is strongly correlated with the wave-derived wind speed in this sample (correlation 0.95; the proxy sits at the lower edge of its window for 60% of the records), and the partial-correlation diagnostic detects little residual linear association between the 0.6–0.9 Hz concentration and inverse wave age after wind speed is accounted for (0.06); this does not exclude a physical wave-age dependence.

## S7. The analytic parametric spectrum

For interpretation, the wind-sea spectrum of Donelan et al. (1985) (DHH) is used as an analytic spectrum in which every frequency can be interrogated:

$$E(f) = \frac{\beta g^2}{(2\pi)^4} f_p^{-1} f^{-4} \exp\left[-(f_p/f)^4\right] \gamma^{r(f)}, \qquad r(f) = \exp\left[-\frac{(f-f_p)^2}{2\sigma^2 f_p^2}\right], \tag{S5}$$

with $\beta = 0.006\,\Omega^{0.55}$, $\gamma = 1.7$ for $\Omega < 1$ and $1.7 + 6\log_{10}\Omega$ above, $\sigma = 0.08(1 + 4\Omega^{-3})$, inverse wave age $\Omega = U_{10}/c_p$ with $c_p$ the phase speed at the peak, and the $\mathrm{sech}^2$ directional lobe of DHH with the Banner extension above $1.6f_p$. A swell partition, where used, is a narrow Gaussian in frequency at its own heading, swept over heights 1–5 m, periods 11–18 s and headings 0–180° relative to the wind sea (Fig. S7).

*Swell and multidirectionality.* Swell can dominate the significant wave height while contributing little to the drift: a swell carrying a fraction 0.77 of the variance contributes 0.04 of the surface Stokes drift at a swell period of 14 s, and at 0.97 of the variance it contributes 0.15 (Fig. S7d). An 8 s swell at the same variance share contributes four times more than an 18 s one. Over the swell sweep the angle between the Stokes vector and the wind sea has quartiles 0.0/0.9/4.1°; only 12.5% of the sweep exceeds 8°, and the largest angle, 29°, is a 5 m, 11 s swell crossing at 120°. The idealised sweep does not determine the origin of the observed median misalignment of 15°. The multidirectional factor $R_{\mathrm{md}}$ over the same sweep has quartiles 0.92/0.98/1.00 with a floor of 0.35 in the crossing corner, against the observed 0.94.

*Spreading.* In these idealised spectra the directional concentration decreases above the peak, and the Stokes-weighted $\langle r_1 \rangle$ is 0.49–0.58 over the tested wave conditions with the integral taken to 3 Hz. Such integrated values depend on the assumed peak frequency, the spectral weighting and the upper integration limit, so they are illustrative and not a like-for-like test of the observed analysis-band $R_{\mathrm{sp}}$ of 0.66 to 1 Hz; Fig. 3 of the main text provides the frequency-resolved comparison with the observations. The DHH lobe was fitted to Lake Ontario data and the Banner extension to a separate fetch-limited data set. The dashed lines in Fig. S7 are the values estimated from the buoy records (Sect. 3.1), shown as targets rather than inputs.

*The DHH normalisation.* Published statements of (S5) differ by a factor $2\pi$ in the normalisation, which is a factor $2\pi$ in spectral level and $\sqrt{2\pi}$ in $H_s$. The normalisation is fixed by conservation of variance between the radian-frequency form in which DHH state their spectrum and the cyclic-frequency form used here, $E(f)\,df = E(\omega)\,d\omega$, which gives the $(2\pi)^{-4}$ form. As a plausibility check, this form integrates at full development to $H_s = 3.02$ m against the Pierson–Moskowitz $0.21U_{10}^2/g = 3.08$ m at $U_{10} = 12\,\mathrm{m\,s^{-1}}$, and its equilibrium level $f^4E$ exceeds the measured level of the recovered fleet at matched wind speed by 1.7×, inside the factor-two spread of the Toba

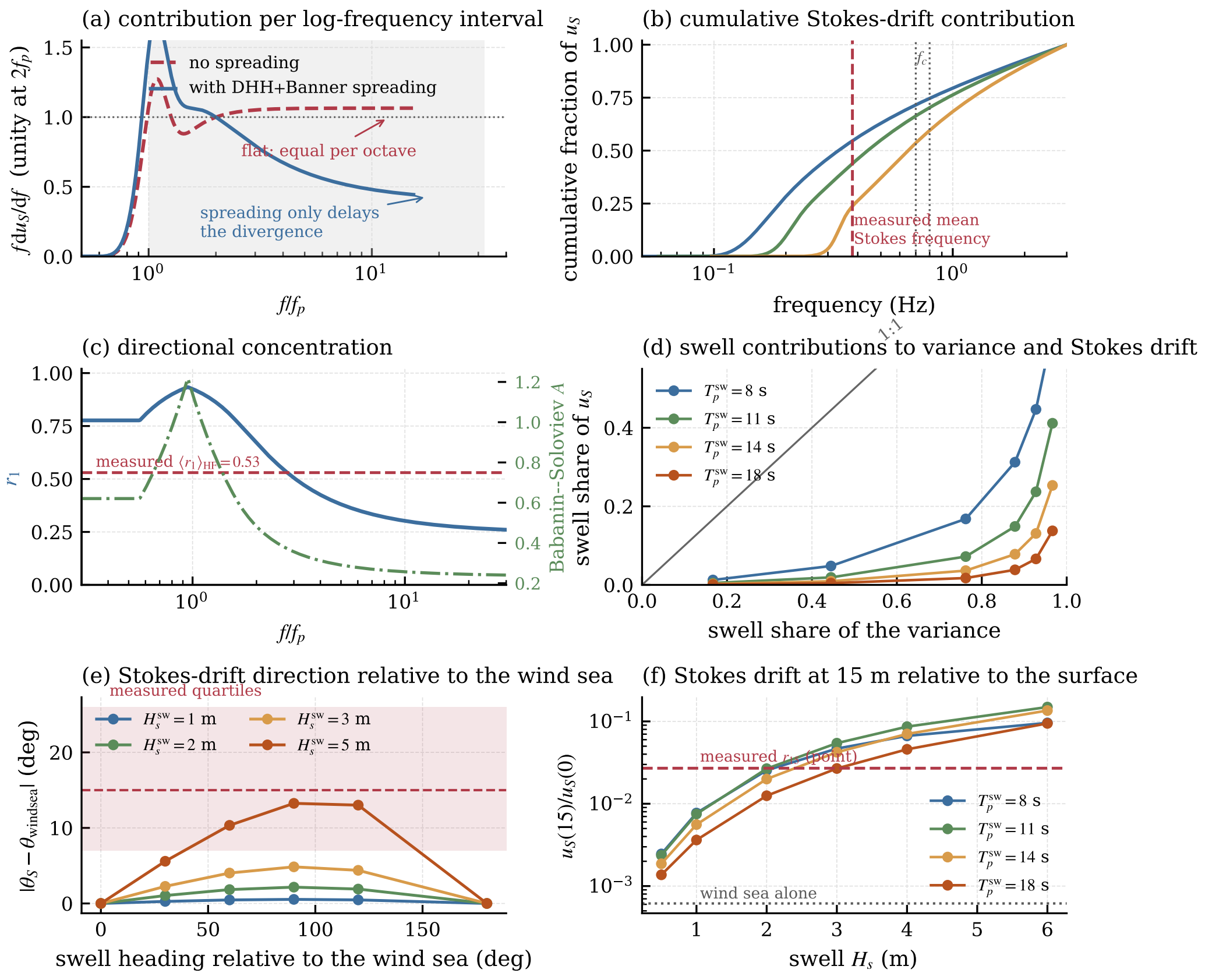


Figure S7: Stokes drift from the analytic spectrum: DHH wind sea with the DHH/Banner $\mathrm{sech}^2$ width, plus a Gaussian swell partition. Dashed lines are the values estimated from the buoy records, shown as targets rather than inputs. (a) The Kenyon integrand times frequency, flat without spreading. (b) Cumulative fraction of $u_s$ against frequency for three wave ages, integral to 3 Hz. (c) First directional moment $r_1$ and the Babanin–Soloviev width against $f/f_p$. (d) Swell's contribution to the variance against its contribution to the surface Stokes drift, by swell period. (e) Angle between the Stokes vector and the wind sea against swell heading. (f) Stokes drift at 15 m depth relative to the surface value against swell height.

constant; the $(2\pi)^{-3}$ form would give $H_s$ 2.5 times too large. Extrapolated to a fully developed sea, identifying its $f^{-4}$ range with Toba's returns $u_* \approx 1.2\,\mathrm{m\,s^{-1}}$ against $0.42\,\mathrm{m\,s^{-1}}$ from a bulk drag law at that wind speed, so DHH's equilibrium level sits about three times above Toba's where DHH was never calibrated. The analytic spectrum therefore fixes the spectral shape only, and the tail level of the main text comes from the buoy-derived friction velocity.

## S8. Parameterised spreading for the telemetered dataset

For the telemetered dataset (tier T1), which archives omnidirectional $S_{\mathrm{acc}}(f)$, the directional distribution is parameterised: direction equal to the buoy-derived wind-sea direction, spreading from DHH with the Banner high-frequency broadening, no fitted constants. This feasibility test uses a separate band-averaged estimator with a 1 Hz cutoff and a Phillips tail with $\alpha_P = 0.0083$, rather than the 0.7 Hz reference estimator of the main text, with the acceleration-spectrum peak as the peak-frequency proxy of the spreading parameterisation; it parameterises the wind-sea directionality while retaining the measured swell contribution, so it is not a telemetry-only reconstruction where no directional information exists. Statistics are evaluated on the 21,006 records of the recovered tier with finite estimates and a directionally resolved speed above $0.02\,\mathrm{m\,s^{-1}}$ (Fig. S8). On them the parameterised vector has a magnitude correlation of 0.90 with the measured one and a median direction difference of 7.9°, but little per-record skill in $r_1$ (correlation 0.14–0.48 by band), and its overestimated directional concentration inflates the magnitude by a median ratio of 1.27: 1.25 on OTC25-MELODI-04, 1.31 on OTC25-MELODI-16, 1.15 on OTC25-MELODI-20 and 1.29 on COI_2025. On the separate open-ocean dataset (S11) the ratio is 1.16 (1.23 and 1.13 on the two buoys). An operational correction is a division by a calibrated ratio, a factor 0.79 for 1.27; dividing by the in-sample median is not a held-out validation, and the deployment-to-deployment spread of 1.15–1.31 together with the 1.16 of the separate open-ocean dataset is the present measure of the calibration's transfer. The resulting calibration is specific to this band-averaged configuration and has not been validated as a correction to the per-frequency 0.7 Hz reference estimator or to a telemetry-only product without measured swell information. On the telemetered dataset the parameterised vector therefore requires a separate-deployment calibration and a quality flag, and it is never a substitute where moments exist.

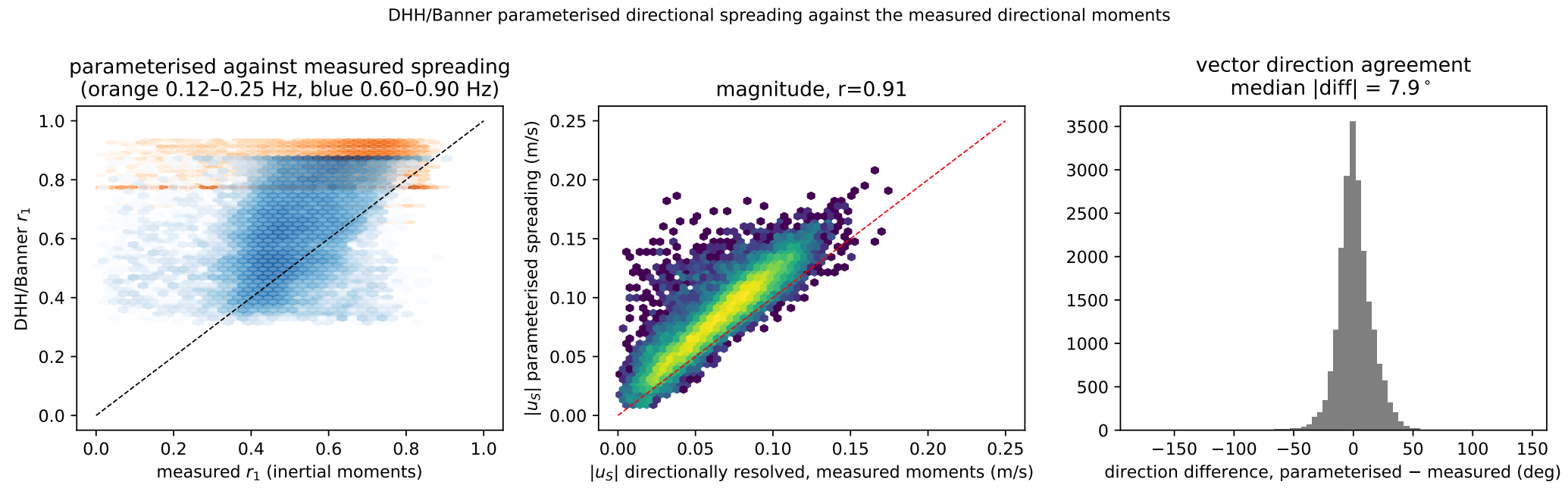


Figure S8: Parameterised directional spreading (band-averaged estimator, 1 Hz cutoff, Phillips tail) against the measured inertial moments on the recovered tier, all records with finite estimates: (a) $r_1$ in the 0.12–0.25 Hz and 0.60–0.90 Hz bands, little per-record skill and an overestimated concentration. (b) Magnitude, about 1.27 times the inertial estimate at the median; the correlation quoted in the text (0.90) is for the 21,006 records with a directionally resolved speed above 0.02 m s$^{-1}$. (c) Vector direction, median difference 7.9°.

## S9. The kinematic draft factor

The draft factor is the ratio of the magnitude of the Stokes-drift vector averaged over the hull draft to its magnitude at the surface, in a purely kinematic calculation on the linear Stokes profile; it is not a measured hull-response coefficient. For a sea-surface floater with a hull of draft $d$, uniform averaging of the leading-order Stokes profile over the draft multiplies the integrand by

$$D(f;d) = \frac{1}{d}\int_{-d}^{0} e^{2kz}\,dz = \frac{1-e^{-2kd}}{2kd}, \tag{S6}$$

which for $d = 6$ cm, the nominal draft of about 5.5 cm rounded, is 0.94 at 0.5 Hz and 0.86 at 0.8 Hz. The kinematic factor $C_s^{\mathrm{kin}} = |\int D\, d\hat{\boldsymbol{u}}_s|/|\int d\hat{\boldsymbol{u}}_s|$ was evaluated over the full reference estimate, resolved part and tail, on a 2,000-record subset of the recovered tier (Fig. S9). For the 6 cm hull it is 0.97 on the resolved band to 0.7 Hz (0.95–0.98, 10–90%), 0.96 on the analysis band to 1 Hz, 0.55 for the tail alone, and 0.82 over the resolved spectrum plus tail (0.65–0.90): 0.72 in calm winds, 0.90 at 10–14 m s$^{-1}$ and 0.92 in strong winds. It depends on the draft as 0.87, 0.82 and 0.75 for 3, 6 and 12 cm. For a constant-direction pure saturation tail the 6 cm kernel averaged over the tail above 0.7 Hz is 0.54 and above 1 Hz 0.41, which is why a factor quoted on the resolved band alone cannot describe the full estimate. In 0.4% of the records, crossing seas with $r_1$ below 0.3, the factor exceeds unity, because

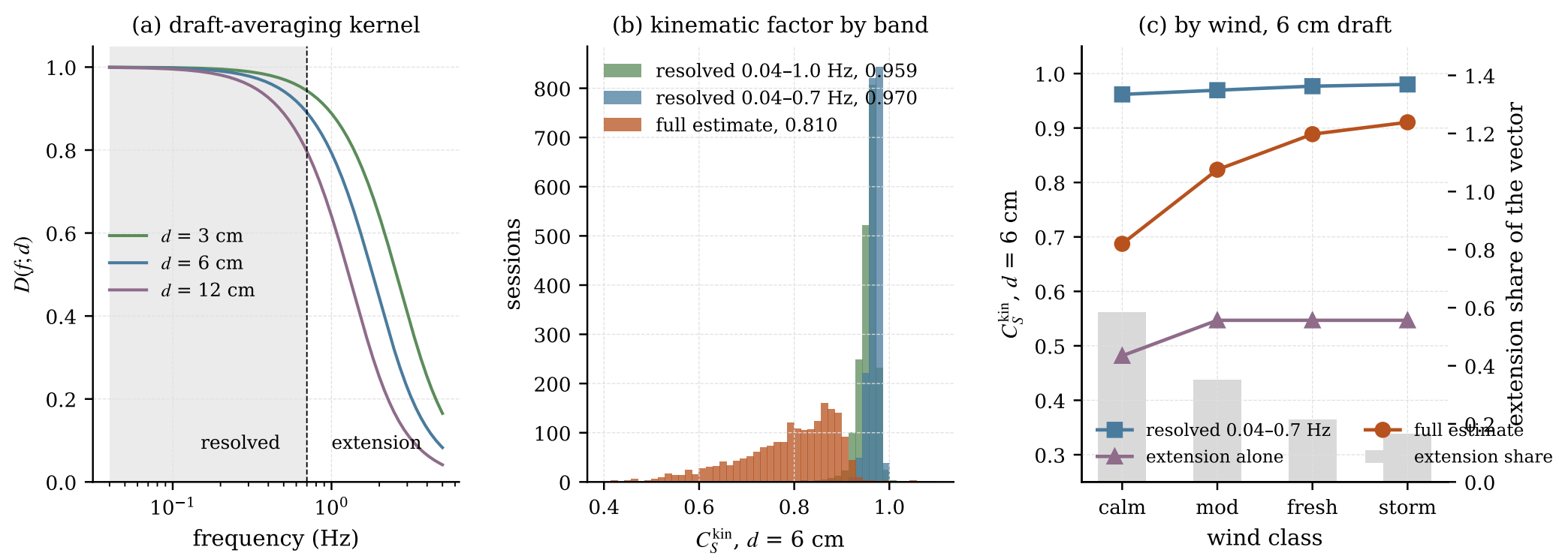


Figure S9: Kinematic draft-averaging factor (2,000-record subset of the recovered tier, reference estimator). (a) Kernel $D(f;d)$ for $d = 3, 6, 12\,\text{cm}$; the shaded range is the resolved band, the tail lies above 0.7 Hz. (b) Distribution of the factor for the 6 cm hull on the analysis band to 1 Hz, on the resolved band to 0.7 Hz and over the full estimate with the tail. (c) The three factors and the tail's contribution to the vector by wind class.

attenuating a cancelling high-frequency component raises the vector. The monochromatic draft correction at the spectral peak gives 0.998 and misses the reduction almost entirely.

These numbers describe uniform draft averaging of the leading-order profile. They are not the hydrodynamic response of a finite floating body: with a hull diameter of 0.025–0.15 of the deep-water wavelength between 0.4 and 1.0 Hz (0.39 at 1.6 Hz), the submergence, buoyancy, drag and added-mass effects treated by Calvert et al. (2021) are not evaluated, and a hull that follows the surface samples the leading-order profile only approximately. No independently validated heave response amplitude operator is available for these deployments, so no hull-response correction is applied to the measured spectrum, and possible response errors remain a limitation of the inferred surface Stokes drift. The estimate of the main text is reported without $D$, as the surface Stokes drift, so that it can be compared with the model's Stokes drift at the sea surface.

## S10. Nonlinear and higher-order effects

The estimator uses the leading-order, second-order-in-amplitude spectral expression for the Stokes drift, evaluated from buoy-derived, quasi-Lagrangian spectra. This section reviews physical effects outside that leading-order spectral estimate. Their magnitudes have not been established for the

present observations, so the literature examples quoted in the paragraphs below are not an uncertainty budget for this dataset.

*Finite amplitude.* The particle orbit stays vertically symmetric to $O((ak)^6)$ (Longuet-Higgins, 1987); this concerns the orbit and does not by itself establish the accuracy of the leading-order drift up to the limiting wave. For random seas the third-order extension of Kenyon's result in the Zakharov–Hamiltonian framework (Stuhlmeier, 2026) introduces a positive difference-harmonic contribution to the mean drift whose importance depends on the spectral shape and on the high-frequency support of the spectrum; in that study's wind-sea example resolved to the breaking wavenumber, it enhances the surface drift substantially while changing the depth-integrated transport by a few per cent. We have not evaluated this contribution for the present directional spectra and assign no numerical correction or uncertainty bound to it.

*Directional spreading in the vector form.* The reductions of Webb and Fox-Kemper (2015) are corrections relative to the unidirectional assumption. In deep water the kernel is direction-independent, so the vector integral over $(a_1, b_1)$ contains the full spreading effect by definition of the first moments. Residual directional error enters through noise in the measured moments (S3), through differential channel response and phase, through the frequency-frame treatment and through the unresolved directional structure of the tail; the exact first-moment identity does not guarantee the last three.

*The measured spectrum is Lagrangian.* The standard spectral expression is formulated in terms of the free-wave spectrum. An Eulerian record carries $O((ak)^2)$ bound-harmonic energy, comparable to the free high-frequency range by $3f_p$ (Janssen, 2009). Surface-following motion suppresses sum-frequency bound components and enhances difference-frequency contributions (Longuet-Higgins, 1986; Herbers and Janssen, 2016; McAllister and van den Bremer, 2019), so the buoy heave spectrum in the wind-sea band is closer to the free-wave spectrum than a wave-staff record. Contributions below 0.04 Hz are excluded by the integration limit, but the residual bound-wave bias within the retained band has not been quantified for these records; no correction is applied. Field comparisons report no significant bias of spectral or directional statistics from buoy motion at ocean steepness (McAllister and van den Bremer, 2020; Herbers et al., 2012), although none of them concerns a hull as small as this one.

*Group-scale motions.* Wave-coherent horizontal velocity of the buoy in the group-frequency band can contain Stokes drift fluctuations (Herbers and Janssen, 2016), Eulerian wave-group motions and platform-related contributions. Under random phases the phase-dependent contributions vanish in an ensemble average (Stuhlmeier, 2026); their residual in an individual 22-min record has not been quantified here, and no record-specific correction is applied. The passing group also pumps an Eulerian counterflow that could cancel the surface Stokes signal at group timescales (Smith, 2006).

*Drift associated with wave breaking.* Breaking converts orbital motion into vortical flow; the spectral estimate contains only the drift of the wave field itself and omits this contribution. Integrated over breaking-crest statistics, field data give a breaking-induced drift $U_{LB} = 0.24\, u_*^{2.2}$ against a spectral Stokes drift $U_S = 0.53\, u_*$ (both in $\mathrm{m\,s^{-1}}$ with $u_*$ in $\mathrm{m\,s^{-1}}$) (Deike et al., 2017; Pizzo et al., 2019): up to 0.3 of the Stokes drift above $U_{10} \approx 15\,\mathrm{m\,s^{-1}}$ and negligible below 5. Separately, steep non-breaking focusing groups produce a local, episodic drift enhancement (Blaser et al., 2026); that result concerns the focusing region and does not establish a wind-speed or friction-velocity scaling for the present records. Neither the breaking contribution nor the focusing enhancement has been evaluated for these observations, and neither is included as an additional term in the leading-order spectral estimate. Whether a 6 cm float samples the associated motions in full is a question of body response that is not answered here.

## S11. The open-ocean telemetry buoys and the high-frequency channel

Two drifters running the firmware of August 2026, TEST_EODYN_001 and 002, drifted west of Biscay (46–47.3°N, 10.1–11.2°W) on 1–3 August 2026 on a 30-min cycle, transmitting the onboard vertical displacement spectrum (0.031–0.992 Hz, 83 bins) with $a_1, b_1$ below 0.5 Hz, the onboard wind and positions. Of 132 records, 125 pass quality control ($H_s > 0.5\,\mathrm{m}$, $r_1 > 0.3$ in 0.3–0.5 Hz, wind available): $H_s = 1.06\,\mathrm{m}$, $f_p = 0.125\,\mathrm{Hz}$, $U_{10} = 7.4\,\mathrm{m\,s^{-1}}$ (5.1–10.7) after the onboard wind was recomputed offline with the retrieval of the recovered tier, drift speed $0.22\,\mathrm{m\,s^{-1}}$. The moments are transmitted below 0.5 Hz only, so the tier cannot check the short-wave level and its resolved band stops at 0.5 Hz.

On 0.04–0.5 Hz the resolved unidirectional and directionally resolved estimates are 0.033 and $0.019\,\mathrm{m\,s^{-1}}$, $R = 0.58$ (0.54 and 0.62 on the two buoys),

the drift-Doppler factor $D_\mathrm{d}$ is 1.08, and the vector lies within 8.8° of the buoy wind direction at the median. The DHH-width fallback of S8 overpredicts $R$ by a factor 1.16 (1.23 and 1.13). Above $1.5f_p$ the moments fall on the recovered-tier curve ($r_1 = 0.69$, 0.70, 0.64 at 1.5–2, 2–3 and $3$–$4f_p$ against 0.70, 0.67, 0.64); at $1$–$1.5f_p$ they are lower (0.50 against 0.75) because the 8 s swell and the wind sea overlap there. The operational SATGLOB forecast $H_s$ is 1.27 times the buoy's at these dates (recovered tier: the hindcast 1.08 times the buoy's), and the WW3 files accessed for these 2026 comparisons contained no Stokes-drift output, so the tier supports the directional moments and not a level comparison.

*The recovered buoy's channel above 0.7 Hz.* Three comparisons bear on the band above the matching frequency. On the 100 Hz accelerometer of the pilot tier the spectral slope of $E(f)$ is −4.3 in both 0.5–0.7 and 0.7–1.0 Hz, on the 3.2 Hz channel of the recovered tier it steepens from −4.8 to −5.6, and the model tail gives −4.7. A Sofar Spotter co-deployed with one recovered buoy provides 279 matched records within 30 km and one hour, median separation 21 km; the median MELODI-to-Spotter energy ratios are 0.93, 1.00, 1.02 and 0.96 in the 0.20–0.30, 0.30–0.40, 0.40–0.51 and 0.51–0.80 Hz bands, and the broad final band constrains neither the spectral shape within it nor the energy at higher frequencies. Emulating the channel's sampling and averaging on the pilot records reproduces a fraction 0.1–0.2 of the steepening; the internal filtering of the attitude unit is not documented. The shortfall above 0.7 Hz is thus consistent with a measurement-response difference of the recovered buoy's channel, but the pilot samples different conditions and the comparisons do not assign it uniquely. The band carries a fraction 0.09 of the buoy's drift resolved to 1 Hz, and the matching frequency of 0.7 Hz keeps it out of the resolved part of the reference estimate.

## S12. Full comparison statistics and glossary of ratios

Table S6 gives the pairwise statistics of the WAVEWATCH III Stokes drift at the sea surface against the buoy reference estimate per deployment and wind class (differences and ratios as model relative to buoy), and Fig. S10 the component and direction comparisons and two storm windows. With $u_b$ and $u_m$ the complex (east +i north) buoy and model vectors and overbars

their sample means, the centred complex vector correlation is

$$r_{\mathrm{v}}^{\mathrm{c}} = \frac{\left|\sum (u_b - \bar{u}_b)(u_m - \bar{u}_m)^*\right|}{\sqrt{\sum |u_b - \bar{u}_b|^2 \sum |u_m - \bar{u}_m|^2}} \tag{S7}$$

(0.92); the uncentred form in the table gives 0.94 and the pooled component correlation 0.93. The mean absolute direction difference is 14° on all records and 11° on the records where both speeds exceed $0.02\,\mathrm{m\,s^{-1}}$. The root-mean-square vector difference is $0.039\,\mathrm{m\,s^{-1}}$, 0.47 of the buoy mean speed, and the slope through the origin of model on buoy speed 1.17. The speed correlation between the buoy estimate and the bulk relation of Ardhuin et al. (2009) (below) is 0.71; a triple collocation of buoy, model and bulk relation is not admissible because the buoy estimate and the bulk relation share the measured sea state.

*Copernicus Marine MFWAM analysis.* The surface Stokes drift of the Copernicus Marine global wave analysis (MFWAM, 1/12°, 3-hourly; E.U. Copernicus Marine Service Information (CMEMS), 2025) was collocated to the same records (bilinear in space, nearest analysis within 90 min; 12,889 records, the Alboran records nearest the coast having no valid bilinear interpolation). Against the buoy-derived estimate $\rho_{\mathrm{MF}} = 1.16$ (median recordwise ratio 1.11), the speed correlation 0.86, the centred vector correlation 0.92, the root-mean-square vector difference $0.042\,\mathrm{m\,s^{-1}}$ and the mean absolute direction difference 15° (12° where both speeds exceed $0.02\,\mathrm{m\,s^{-1}}$); by wind class $\rho_{\mathrm{MF}}$ is 0.85, 1.17, 1.39 and 1.62 against $\rho_{\mathrm{WW3}} = 0.82$, 1.13, 1.38 and 1.61 on its own collocated sample, and per deployment 1.15, 1.19 and 0.90 against 1.13, 1.15 and 0.90 (Fig. S11, Table S7). On the 12,865 records where both products are available, $\rho_{\mathrm{WW3}} = 1.13$ and $\rho_{\mathrm{MF}} = 1.16$, the speed correlations 0.87 and 0.86, the centred vector correlations 0.92 and 0.92, and the mean absolute direction differences 13° and 15°. MFWAM and WAVEWATCH III agree with each other to a MFWAM-to-WW3 ratio of mean speeds of 1.02, a vector correlation of 0.99 and a direction difference of 7°. Sampling intervals of the wind-class ratios of mean speeds (Fig. 6c, f of the main text), from 2,000 resamplings of the 3-day blocks within each deployment with the estimator constants held fixed, are 0.79–0.86, 1.10–1.15, 1.34–1.41 and 1.49–1.72 for WAVEWATCH III (83, 91, 57 and 18 occupied blocks) and 0.81–0.90, 1.14–1.19, 1.35–1.42 and 1.50–1.73 for MFWAM; they describe sampling uncertainty conditional on the estimator and exclude tail and instrument uncertainty.

Table S6: WAVEWATCH III Stokes drift at the sea surface against the buoy reference estimate, per deployment and wind class: number of records, mean speeds, normalised bias of the model mean speed relative to the buoy, root-mean-square vector difference, speed correlation, slope through the origin of model on buoy speed, uncentred complex vector correlation $r_\mathrm{v}$ and mean absolute direction difference (all records, no minimum speed threshold).

| subset | $N$ | $\overline{\lvert u_s \rvert}$ | $\overline{\lvert u_s^{\mathrm{WW3}} \rvert}$ | bias | RMSE | $r$ | slope | $r_\mathrm{v}$ | $\Delta\theta$ |
|---|---|---|---|---|---|---|---|---|---|
| | | $\mathrm{m\,s^{-1}}$ | $\mathrm{m\,s^{-1}}$ | % | $\mathrm{m\,s^{-1}}$ | | | | deg |
| all | 13,115 | 0.082 | 0.093 | +13 | 0.039 | 0.87 | 1.17 | 0.94 | 14 |
| OTC25-MELODI-04 | 9,105 | 0.084 | 0.095 | +13 | 0.040 | 0.88 | 1.17 | 0.94 | 13 |
| OTC25-MELODI-16 | 3,384 | 0.084 | 0.097 | +15 | 0.038 | 0.83 | 1.18 | 0.95 | 12 |
| OTC25-MELODI-20 | 626 | 0.044 | 0.040 | −10 | 0.026 | 0.84 | 0.91 | 0.89 | 35 |
| $U_{10} < 6$ | 5,272 | 0.053 | 0.044 | −18 | 0.022 | 0.80 | 0.84 | 0.92 | 20 |
| 6–10 | 5,357 | 0.092 | 0.104 | +13 | 0.032 | 0.66 | 1.13 | 0.96 | 9 |
| 10–14 | 2,281 | 0.122 | 0.168 | +38 | 0.065 | 0.30 | 1.34 | 0.95 | 11 |
| $> 14$ | 205 | 0.142 | 0.228 | +61 | 0.108 | 0.30 | 1.49 | 0.93 | 12 |

Table S7: As Table S6, for the Copernicus Marine MFWAM analysis ($N = 12{,}889$ records with a collocated field).

| subset | $N$ | $\overline{\lvert u_s \rvert}$ m s$^{-1}$ | $\overline{\lvert u_s^{\mathrm{MFWAM}} \rvert}$ m s$^{-1}$ | bias % | RMSE m s$^{-1}$ | $r$ | slope | $r_{\mathrm{v}}$ | $\Delta\theta$ deg |
|---|---|---|---|---|---|---|---|---|---|
| all | 12,889 | 0.083 | 0.096 | +16 | 0.042 | 0.86 | 1.19 | 0.93 | 15 |
| OTC25-MELODI-04 | 9,129 | 0.084 | 0.097 | +15 | 0.043 | 0.87 | 1.19 | 0.93 | 15 |
| OTC25-MELODI-16 | 3,384 | 0.084 | 0.100 | +19 | 0.040 | 0.83 | 1.21 | 0.95 | 12 |
| OTC25-MELODI-20 | 376 | 0.052 | 0.047 | −10 | 0.027 | 0.85 | 0.91 | 0.90 | 31 |
| $U_{10} < 6$ | 5,022 | 0.054 | 0.046 | −15 | 0.024 | 0.76 | 0.87 | 0.90 | 22 |
| 6–10 | 5,376 | 0.092 | 0.107 | +17 | 0.036 | 0.62 | 1.16 | 0.95 | 10 |
| 10–14 | 2,286 | 0.122 | 0.170 | +39 | 0.067 | 0.29 | 1.34 | 0.95 | 11 |
| $> 14$ | 205 | 0.142 | 0.229 | +62 | 0.110 | 0.27 | 1.50 | 0.92 | 12 |

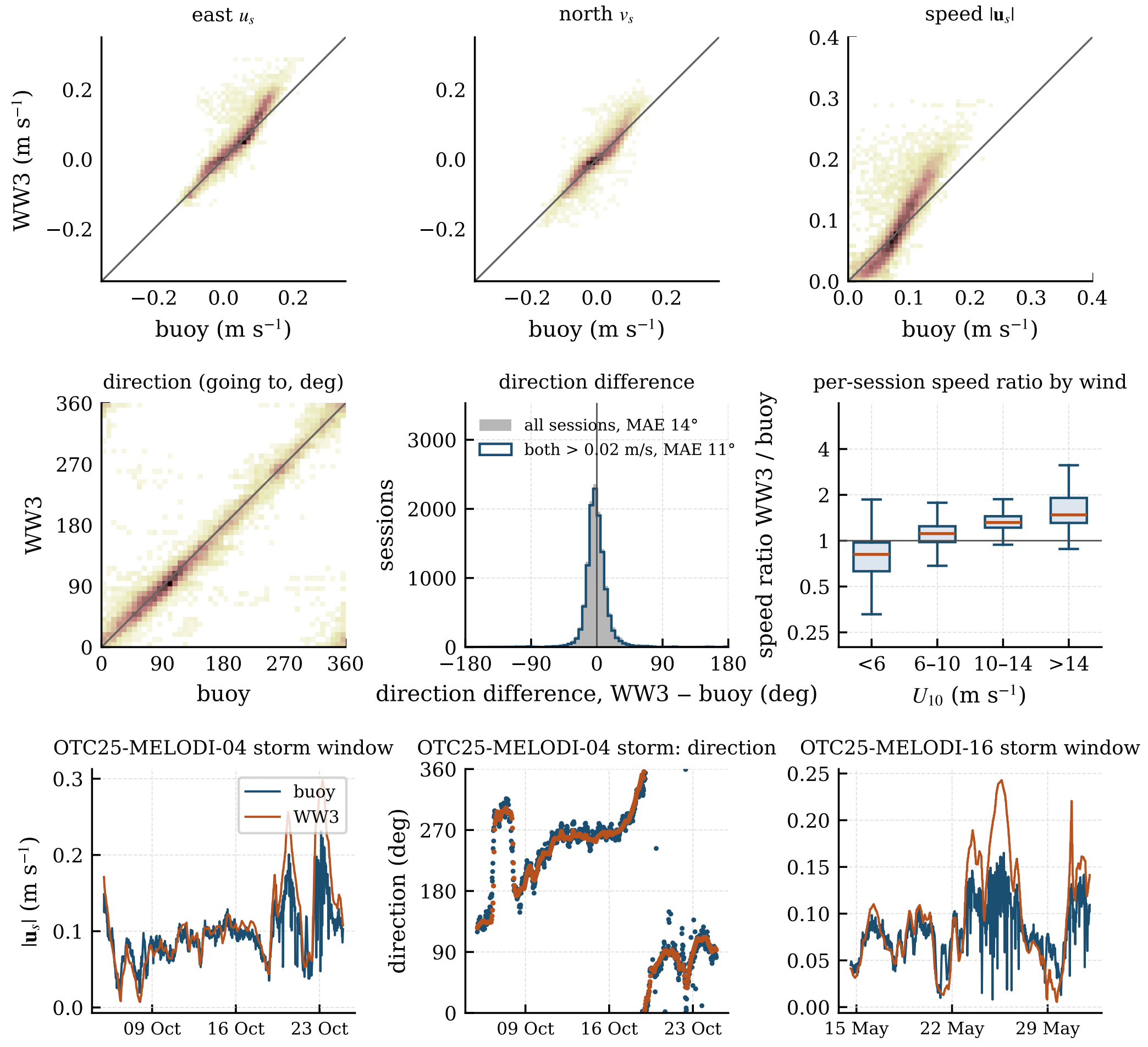


Figure S10: WAVEWATCH III Stokes drift at the sea surface against the buoy reference estimate (buoy on the horizontal axes), $N = 13{,}115$ records: component and speed density scatter (top), direction heatmap, direction-difference distribution (model minus buoy) and per-record model-to-buoy speed ratio by wind class (middle), storm-window time series (bottom).

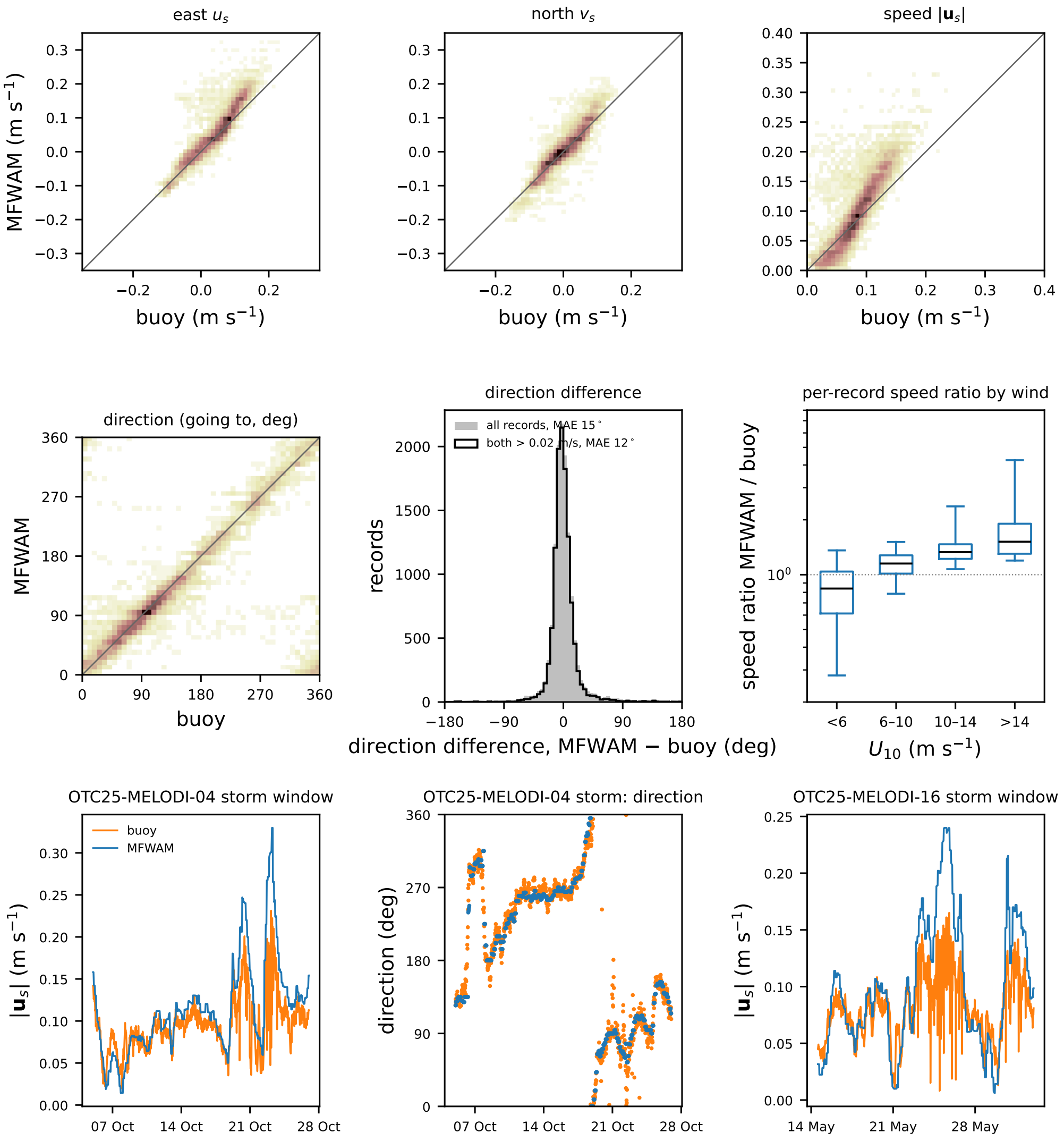


Figure S11: Copernicus Marine MFWAM surface Stokes drift against the buoy reference estimate (buoy on the horizontal axes), $N$ = 12,889 records, in the layout of Fig. S10: component and speed density scatter (top), direction heatmap, direction-difference distribution (model minus buoy) and per-record model-to-buoy speed ratio by wind class (middle), storm-window time series (bottom).

On the analysis band to 1 Hz without tail, the median recordwise ratio of the model vector to the directionally resolved buoy estimate is 1.44, increasing from 1.37 below 6 m s$^{-1}$ to 1.65 above 14 m s$^{-1}$, whereas the model vector is 0.87 of the unidirectional-equivalent buoy estimate. Record by record the first ratio equals the second divided by $R$; the separately reported medians need not satisfy that identity. Per deployment $\rho_{\mathrm{WW3}}$ is 1.13, 1.15 and 0.90 (Table S6). For the station comparison of Sect. 3.2, widening the match from 60 km and 90 min to 120 km and 180 min changes the two directional reductions by at most 0.01 and the amplitude ratios by at most 0.03 (30–120 km and 45–180 min tested).

Figure S12 shows, without ranking, the bulk relation of Ardhuin et al. (2009) evaluated with the buoy's own wind speed and significant wave height, $U_{ss} = 5.0 \times 10^{-4}\,[1.25 - 0.25\,(0.5/f_c)^{1.3}]\,U_{10}\min(U_{10}, 14.5) + 0.025\,(H_s - 0.4)$ (their Eq. 7, $f_c = 2$ Hz; the factor is 1.21, against 1.25 for $f_c \to \infty$ and 1.09 for $f_c = 0.7$ Hz), next to the reference estimate and the collocated WAVEWATCH III drift. On the 13,139 records the bulk relation has the same mean as the reference estimate (ratio of means 1.00, median recordwise ratio 0.85, correlation 0.71), but the agreement is not uniform in wind: the ratio of means is 0.69 below 6 m s$^{-1}$, 0.92 at 6–10, 1.35 at 10–14 and 1.94 above 14 m s$^{-1}$, where the bulk relation also exceeds WAVEWATCH III (1.21). The relation was fitted by Ardhuin et al. (2009) to model output with a different wind product and spectral tail, and it shares the buoy's wind speed and wave height with the reference estimate, so the comparison is descriptive.

*Model products.* Table S8 summarises the configuration of the three model products. The gridded WW3 fields were produced with WW3 version 7.00 and the station spectra with version 7.14 (file attributes), both on 36 frequencies from 0.034 to 0.95 Hz and 24 directions. The distributed WW3 Stokes-drift diagnostic integrates the prognostic spectrum over these frequencies; the analytical tail term is disabled in the output code of the public releases 6.07 to 7.14 (The WAVEWATCH III Development Group, 2024). MFWAM integrates over 30 frequencies from 0.035 to 0.58 Hz and 24 directions and, in the public ECWAM routine on which the operational code is based, adds the analytic contribution of an $f^{-5}$ tail matched to the spectral level at 0.58 Hz (ECMWF, 2024). The gridded WW3 fields were interpolated bilinearly in space and linearly in time between the 3-hourly outputs to the 10-min buoy positions and matched to the record times within 40 min.

Several ratios of similar appearance arise in the main text. Table S9 de-

Table S8: Model products compared with the buoy-derived estimates.

| | WW3 gridded fields | WW3 station spectra | MFWAM analysis |
|---|---|---|---|
| product | GLOBMULTI_ERA5_GLOBCUR_01 (Ifremer/LOPS hindcast) | GLOB-30M_LOPS_2025 (Ifremer/LOPS hindcast) | GLOBAL_ANALYSIS-FORECAST_WAV_001_027 (Copernicus Marine) |
| model, version | WW3 7.00 | WW3 7.14 | MFWAM (Météo-France) |
| grid | 10′ North-East Atlantic, 0.5° global fallback | 0.5° global, 140 output stations | 1/12° global |
| forcing | ERA5 winds, GLOBCURRENT currents | ERA5 winds; Alday and Ardhuin (2023) | operational analysis |
| output | surface Stokes drift vector, 3-hourly | directional spectra, 3-hourly | surface Stokes drift vector, 3-hourly |
| spectral support | 36 f, 0.034–0.95 Hz, 24 dir; no tail | 36 f, 0.034–0.95 Hz, 24 dir | 30 f, 0.035–0.58 Hz, 24 dir; $f^{-5}$ tail |
| collocation | bilinear in space, linear in time, $\leq$ 40 min | nearest station $\leq$ 60 km, $\leq$ 90 min | bilinear in space, nearest analysis $\leq$ 90 min |
| records | 13,115 | 4,610 (4,383 buoy 04, 227 buoy 16) | 12,889 |

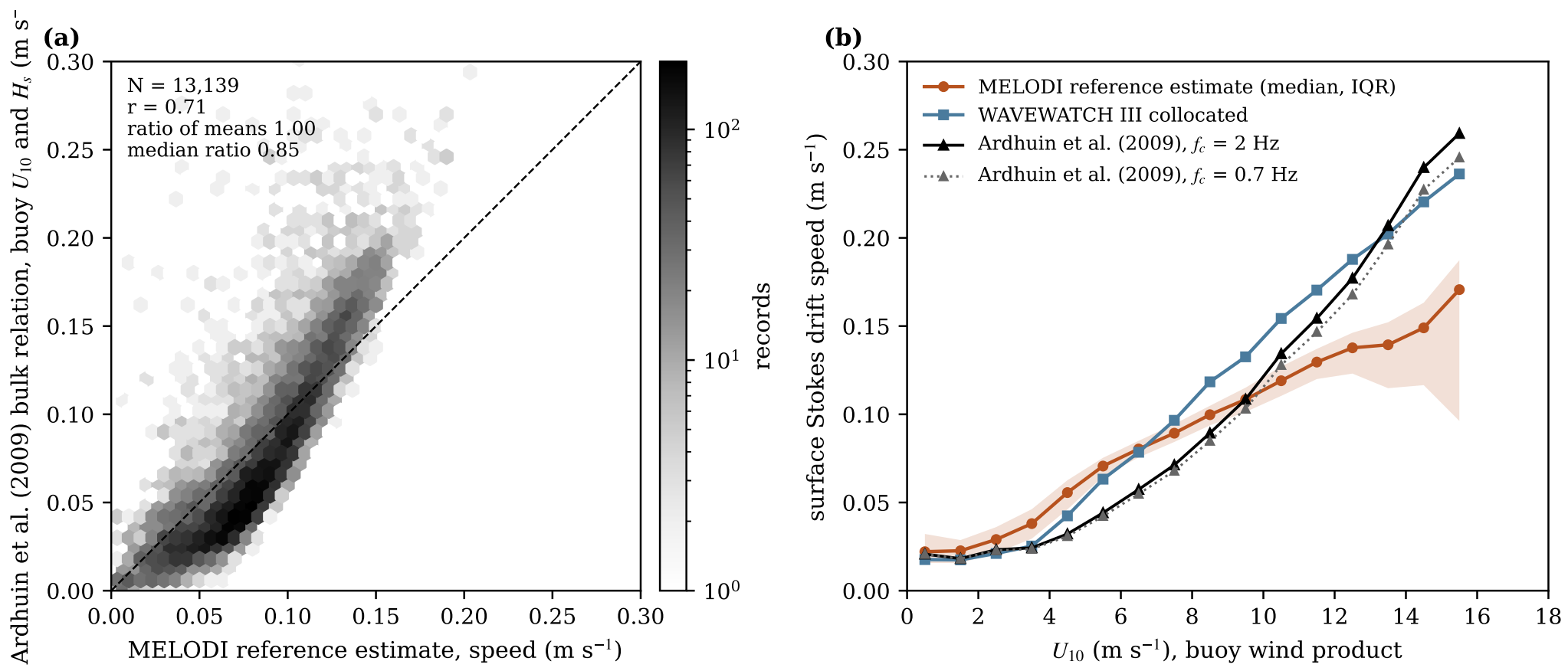


Figure S12: The bulk relation of Ardhuin et al. (2009) evaluated with the buoy's wind speed and wave height ($f_c = 2$ Hz, the cutoff of the empirical relation, distinct from the 0.7 Hz matching frequency of the reference estimate) against the reference estimate ($N = 13{,}139$). (a) Density scatter of the two speeds with the 1:1 line. (b) Medians in 1-m s$^{-1}$ wind bins of the reference estimate (with interquartile range), the collocated WAVEWATCH III drift and the bulk relation for $f_c = 2$ and 0.7 Hz.

fines each, with the buoy quantity, the statistic and the population. The model value is always the WAVEWATCH III Stokes drift at the sea surface, the integral of the prognostic spectrum to 0.95 Hz with the model's own directional reduction and no tail contribution, except where the station spectra or MFWAM are named. Buoy values include the drift-Doppler correction unless stated. A uniform spectral-level factor $c$ between two spectra of equal directional shape gives a Stokes ratio $c$ and a wave-height ratio $\sqrt{c}$, so the ratio of mean speeds and the ratio of wave heights are not expected to coincide, and a wave-height ratio of 1.08 corresponds to a Stokes ratio of 1.16.

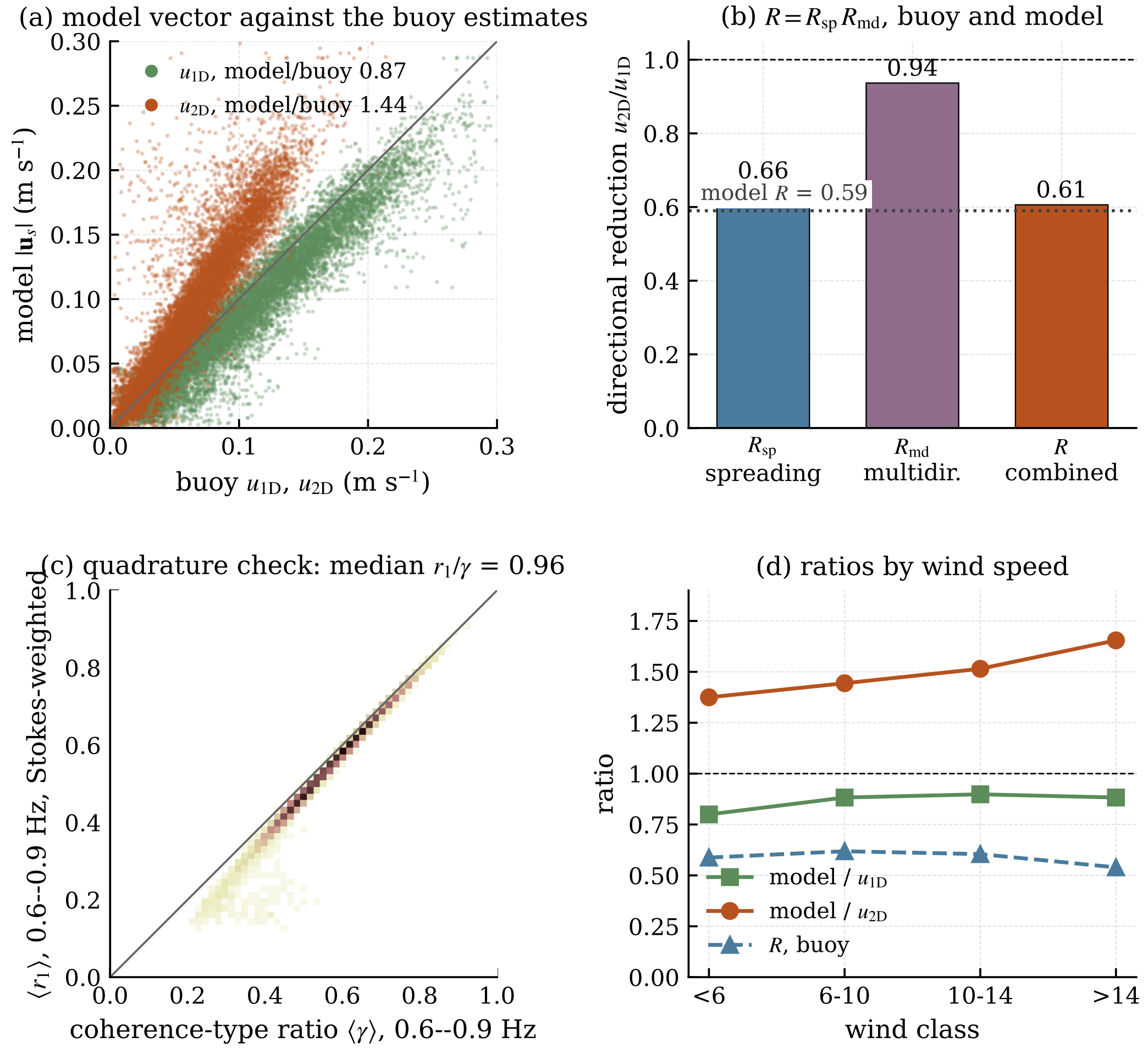


Figure S13: Decomposition of the analysis-band model-to-buoy ratio ($N = 13{,}115$, analysis band to 1 Hz). (a) Model vector against the unidirectional-equivalent buoy estimate $u_{1D}$ (model-to-buoy ratio 0.87) and the directionally resolved estimate $u_{2D}$ (1.44), buoy on the horizontal axis. (b) The buoy's reduction $R = R_{sp}R_{md}$ is dominated by spreading within frequencies; the dotted line is the model's own $R = 0.59$ on its spectra along the tracks. (c) Stokes-weighted $\langle r_1 \rangle$ in 0.60–0.90 Hz against the coherence-type ratio $\gamma$ formed from the same cross-spectra, a quadrature check. (d) The two ratios and the buoy's $R$ by wind class.

Table S9: Model-to-buoy amplitude ratios used in the main text.

| value | buoy quantity in the denominator | statistic | population |
|---|---|---|---|
| 1.13 | reference estimate (matching frequency 0.7 Hz, tail included) | $\rho_\mathrm{m}$ (mean speeds) | fields, $N = 13{,}115$ |
| 1.19 | as above, without the drift-Doppler correction | $\rho_\mathrm{m}$ (mean speeds) | fields |
| 1.49 | reference estimate truncated at 0.95 Hz, approximately matching the last WW3 bin centre | $\rho_\mathrm{m}$ (mean speeds) | fields |
| 1.16 | reference estimate, Copernicus Marine MFWAM analysis in the numerator | $\rho_\mathrm{m}$ (mean speeds) | MFWAM fields, $N = 12{,}889$ |
| 1.17 | reference estimate | slope through the origin of model on buoy speed | fields |
| 0.82 → 1.61 | reference estimate by wind class ($< 6$ to $> 14\,\mathrm{m\,s^{-1}}$) | $\rho_\mathrm{m}$ (mean speeds) | fields |
| 1.44 | directionally resolved, analysis band 0.04–1.0 Hz, no tail | median of per-record ratios | fields |
| 1.37 → 1.65 | as above, by wind class | median per class | fields |
| 0.87 | unidirectional-equivalent, analysis band | median of per-record ratios | fields |
| 0.61 | $R = u_{2\mathrm{D}}/u_{1\mathrm{D}}$, buoy only, analysis band | median | fields |
| 1.24, 1.31 | directionally resolved and unidirectional-equivalent, analysis band, station spectra in the numerator | median of per-record ratios | stations, $N = 4{,}610$ |
| 1.08 | significant wave height, station spectra in the numerator | median of per-record ratios | stations |
| 1.20, 1.15, 1.96 | directionally resolved, intrinsic bands $< 0.5$, 0.5–0.7, 0.7–1.0 Hz, station spectra in the numerator | median of per-record ratios | stations |
| 1.17 | fields against the station spectra integrated to 0.95 Hz (no tail, different configuration and support; model-to-model) | median | stations |


\end{document}